\documentclass[10pt,journal,compsoc]{IEEEtran}
\ifCLASSOPTIONcompsoc
  \usepackage[nocompress]{cite}
\else
  \usepackage{cite}
\fi
\ifCLASSINFOpdf
  \usepackage[pdftex]{graphicx}
\else
\fi
\graphicspath{{figures/}{pictures/}{images/}{figs/}{./}} % where to search for the images

\usepackage{amsmath}
\usepackage{wrapfig}
\usepackage{amsfonts}
\usepackage{enumitem}
\usepackage{amssymb}
\usepackage{hyperref}
\usepackage[mathscr]{euscript}
\usepackage{xcolor}

\newcommand{\RNum}[1]{\uppercase\expandafter{\romannumeral #1\relax}}
\newcommand{\figref}[1]{Fig.~\ref{#1}}
\newcommand{\tabref}[1]{Table~\ref{#1}}

\newcommand{\secref}[1]{Section~\ref{#1}}

\newcommand{\xwwedit}[1]{#1}
\newcommand{\comments}[1]{}
\newlength\savedwidth

\usepackage{microtype}                 % use micro-typography (slightly more compact, better to read)
\PassOptionsToPackage{warn}{textcomp}  % to address font issues with \textrightarrow
\usepackage{textcomp}                  % use better special symbols
\usepackage{mathptmx}                  % use matching math font
\usepackage{times}                     % we use Times as the main font
\usepackage{cite}                      % needed to automatically sort the references
\usepackage{tabu}                      % only used for the table example
\usepackage{booktabs}                  % only used for the table example

\begin{document}
\title{Computational Design of Skinned Quad-Robots}

\author{Xudong Feng,
		Jiafeng Liu,
        Huamin Wang,~\IEEEmembership{Member,~IEEE}
        Yin~Yang,~\IEEEmembership{Member,~IEEE}
        Hujun~Bao,
        Bernd~Bickel,
        and~Weiwei~Xu,~\IEEEmembership{Member,~IEEE}% <-this % stops a space
\IEEEcompsocitemizethanks{\IEEEcompsocthanksitem Xudong Feng, Jiafeng Liu, Hujun Bao and Weiwei Xu are with State Key Lab of CAD\& CG, Department of Computer Science, Zhejiang university, Zhejiang 310058, China. Weiwei Xu is the corresponding author. \protect\\
E-mail:\{bao, xww\}@cad.zju.edu.cn
\IEEEcompsocthanksitem Huamin Wang is with Department of Computer Science and Engineering, Ohio
State University. E-mail:whmin@cse.ohio-state.edu.
\IEEEcompsocthanksitem Yin Yang are with the Department of Electrical and Computer
Engineering, University of New Mexico. E-mail: yangy@unm.edu.
\IEEEcompsocthanksitem Bernd Bickel are with Institute of Science and Technology Austria, Vienna, Austria. E-mail: bernd.bickel@gmail.com.}
% <-this % stops a space
}



\markboth{August~2019}%
{Feng \MakeLowercase{\textit{et al.}}: Computational Design of Skinned Quad-Robots}



\IEEEtitleabstractindextext{%
\begin{abstract}
We present a computational design system that assists users to model, optimize, and fabricate quad-robots with soft skins.Our system addresses the challenging task of predicting their physical behavior by fully integrating the multibody dynamics of the mechanical skeleton and the elastic behavior of the soft skin. The developed motion control strategy uses an alternating optimization scheme to avoid expensive full space time-optimization, interleaving space-time optimization for the skeleton and frame-by-frame optimization for the full dynamics. The output are motor torques to drive the robot to achieve a user prescribed motion trajectory.We also provide a collection of convenient engineering tools and empirical manufacturing guidance to support the fabrication of the designed quad-robot. We validate the feasibility of designs generated with our system through physics simulations and with a physically-fabricated prototype.
\end{abstract}

}

\maketitle

% is not selected <OR> if conference mode is selected - because compsoc
% conference papers position the abstract like regular (non-compsoc)
\IEEEdisplaynontitleabstractindextext

% \fi
\IEEEpeerreviewmaketitle

\IEEEraisesectionheading{\section{Introduction}\label{sec:introduction}}
\IEEEPARstart{T}he design and construction of robots that can execute compelling
motions is a challenging task. It requires careful geometric planning of robotic mechanisms and professional knowledge of the kinematic and dynamic behavior of the robot. Embedding such knowledge into procedures of computational robot design~\cite{leger:1999:thesis,Desai2017} in conjunction with rapid prototyping techniques, such as 3D printing technology, bears tremendous potential to accelerate the construction of personalized robots. For instance, Megaro et al.~\cite{Megaro:2015:IDR} used a kinematic optimization algorithm for the design of multilegged robots consisting of rigid links. However, real-world creatures are not merely rigid skeleton rigs. The muscle and flesh surrounding the skeleton provides diverse morphologies, enriched expressivity. For instance, people wearing a prosthetic limb often prefer a highly realistic rubbery artificial arm over a more functional mechanical one~\cite{gallagher2001adjustment}. Moreover, skins and muscles might be essential for facilitating challenging tasks, such as reproducing compliant grasping of a hand or swimming motions of a fish~\cite{majidi2014soft,wang2008micro}.

Different to the simulation of the robots with rigid links only, the influence of the soft body on the control torques at actuators and on the contact forces at the ground should be simulated and fully taken care of to judge the plausibility of a designed soft robot. This imposes a significant computational challenge for the motion design and fabrication of skinned robots. In this paper, our goal is to address this problem by integrating both dynamic simulation and kinematic optimization into the motion design of quad-robot systems, with soft skins attached as their organic embodiments. Its kernel is an optimization problem that integrates both user-provided kinematic preference and physical constraints of the robot to obtain a dynamically feasible motion plan. The primary physical constraint is the dynamic viability of the skeleton trajectory when a soft skin is attached. The dynamics of the robot is formulated as two-way coupled subsystems of the rigid multibody system and the deformable skin. In addition, our system also incorporates motor constraints and a stability constraint. The motor constraints ensure that the calculated joint torques are within the physical limits of the installed servo motors, whereas the stability constraint requires the center of projection (COP) of the robot structure to fall inside the supporting polygon. As a result, given the surface mesh and desired kinematic trajectories of the robot, our pipeline generates a  physically-valid motion plan that can be realized under the given hardware constraints. The robot itself can be fabricated using rapid prototyping technology.

\begin{figure*}[t]
\centering
\includegraphics[width=\linewidth]{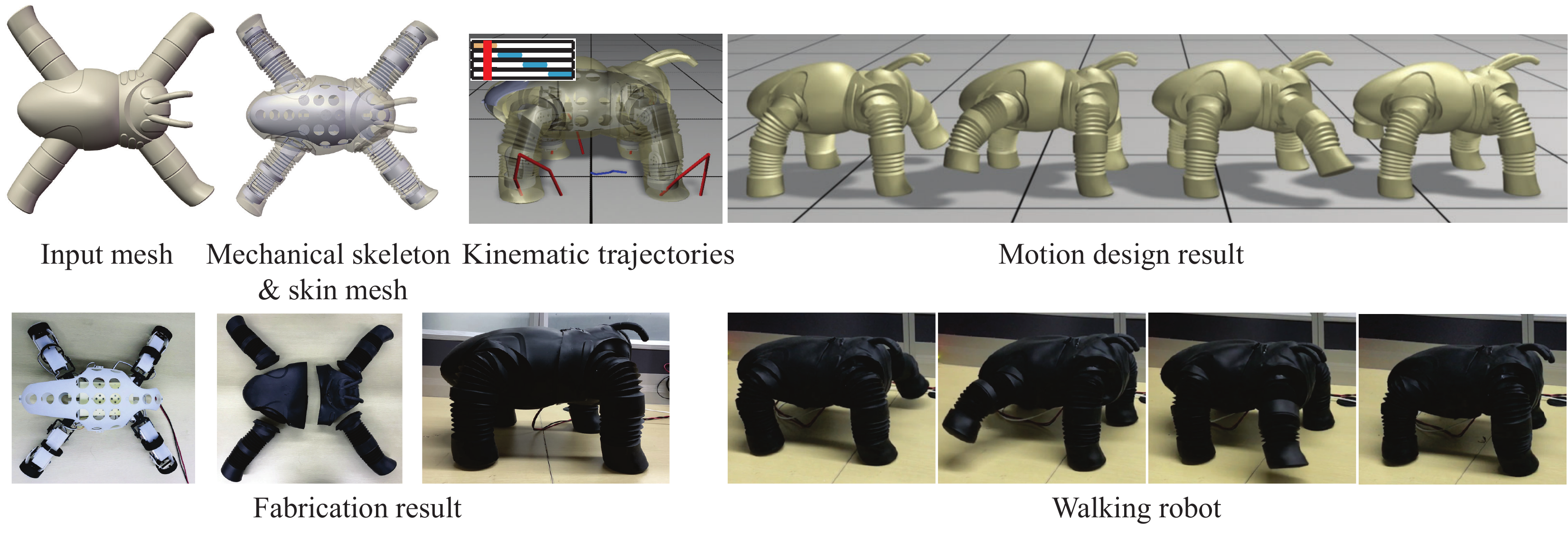}
\caption{We propose a computational fabrication system for designing and fabricating skinned quad-robots. Given an input mesh representing the shape of a quad-robot such as this beetle-like robot, we design its mechanical skeleton with motors to drive its locomotion. The motion plan is generated by an optimization algorithm with kinematic trajectories as the user input. The trajectories consist of the foot swing trajectories (red curves), center of mass (COM) trajectory (blue curve) and foot contact plan (yellow bars) input by the user. Our optimization fully takes account of all the physical and dynamical constraints. By fabricating this robot design, it can be verified that our algorithm is able to generate plausible and physically feasible motion plans for quad-robots, and the simulated results well match the physical experimental results. Note that we only render the transparent input surface without thickness to demonstrate the surface-structure coupling geometry.}
\label{fig:teaser}
\end{figure*}

While space-time optimization is widely used in long-horizon motion design problem, it is computationally prohibitive if directly applied to our case of two-way coupled system, mostly due to the large number of DOFs of the soft skin mesh coupled with the skeleton and various physcial constraints. To this end, we propose an alternating optimization algorithm of two optimization steps to mitigate the computational cost: (1) Space-time optimization with repect to the DOFs of the rigid skeleton while assuming the deformation of the soft skin remains the same as the previous iteration. \xwwedit{(2) Frame-by-frame optimization as in~\cite{Tan:2012:SBL:2185520.2185522} to further optimize the motion plan obtained in step 1 according to all the physical constraints at each frame. To this end, we non-trivially extend the spring-based control force formulation in ~\cite{Tan:2012:SBL:2185520.2185522} to handle the full simulation of the two-way coupled rigid skeleton and soft skin. Our solver can efficiently handle the Lagrange multipliers introduced by the coupling constraints and the collisions between the skeleton and skin.} With a proper initialization, these two steps are alternatively executed until the convergence. In the space-time optimization step, the influence of the soft skin to the rigid skeleton is treated as known quantity and simplified to be the coupling forces and the influence of the center of mass (COM) at each link in the skeleton due to the skin deformation. This setting makes the space-time optimization computationally efficient through decoupling the skin mesh DOFs, which is critical to achieve global effect in the motion design. The skin deformation is updated after each frame-by-frame optimization.

To ease fabrication, we provide a convenient workflow with tailored engineering tools and empirical fabrication guidance embedded in a standard CAD software, empowering regular users to design quad-robots. The modular design scheme allows the user to quickly start from a design template of the mechanical skeleton in \texttt{SolidWorks} and adapt it to the body shape. The rigid skeleton is fabricated via 3D printing, and the skin is separately fabricated using injection molding by pieces. We tested the optimization algorithm on skinned quad-robots with varying body-to-leg ratios and different mechanical skeletons. Both physical and numerical experiments show that the proposed algorithm is an effective means of obtaining physically valid motions of the skinned quad-robots.

\section{Related Work}\label{sec:related_work}

\noindent\textbf{Computational fabrication} aims at designing and creating physical artifacts with the help of computational methods. A large class of methods addresses inverse design problems by incorporating fabrication limitations in geometric design algorithms via constrained optimization or the integration of fast simulation techniques~\cite{Chen:2018:FUB,Dai:2018:SVP}. This line of research enables the design of objects with a wide range of controllable physical and mechanical properties, such as appearance~\cite{Lan:2013:BAF,Chen:2013:SRM,Brunton:2018:PSV,Sakurai:2018:FRD}, deformation~\cite{Bickel:2010:DFM,Skouras:2013:CDA,Panetta:2015:ETA,Chen2017}, articulation~\cite{Lau:2011:CFM,Cali:2012:NAM,Bacher:2012:FAC}, and mechanical motion~\cite{Coros:2013:CDM,Ceylan:2013:DFM,Zhang:2017:FRM,Megaro2017,Geilinger:2018:SOD}. Some existing contributions also investigated how to instantiate virtual characters as 3D-printable physical entities like mechanical robots ~\cite{Coros:2013:CDM,Megaro:2015:IDR}. Yet, these methods merely focus on robots consisting of rigid links and basic balancing constraints and/or velocity limits.

%our work extends to the design of personal robots with soft skin, which try to mimic real creature in daily life.

Bickel et al.~\cite{Bickel:2012:PFC} proposed a process for designing synthetic skin and actuation parameters for animatronic characters that mimic facial expressions of a given subject. Skouras et al.~\cite{Skouras:2013:CDA} optimized the internal material
distribution so that the resulting character exhibits the desired deformation behavior. Focusing on  actuation, Bern et al.~\cite{Bern2017} computed the layout of winch-tendon networks to animate plush toys, and Ma et al.~\cite{Ma2017} optimized the chamber structure and material distribution for designing soft pneumatic objects.

Our work shares some of these goals but takes a significantly different approach. Instead of relying on a rigid or quasi-static underlying simulation of the skin deformation in our case, the influence of the soft skin on the embedded moving mechanical skeleton makes us face a dynamic two-way coupled multibody-elastic problem. It is much more challenging to  accurately simulate and optimize due to the drastically increased system complexity, nonlinearity and discontinuity, i.e., due to the complimentary constraints enforced at the contact vertices.

%	 Additionally, joint torque limits should also be considered to check the feasibility of a user design.

\vspace{5 pt}
\noindent\textbf{Physics-based character motion generation} has vast applications in both graphics and robotics. Algorithmic approaches include leveraging space-time optimization with necessary physical constraints and developing controllers to drive forward simulations.

%\begin{figure*}[t]
%  \centering
%  \includegraphics[width=\textwidth]{figs/flowchart/flowchart.pdf}
%  \caption{A designed skinned beetle-like robot. Given an input mesh representing the shape of a quad-robot, we design its mechanical skeleton with motors to drive its motion. The motion plan is generated by an optimization algorithm with kinematic trajectories input by the user and dynamics constraints of the designed robot, where the trajectories consist of the foot swing trajectories in red, center of mass (COM) trajectory in blue and foot contact plan (top-left) input by the user. Bottom right: the walking motion of the fabricated robot driven by the motion plan. Note that we only render transparent input surface without thickness to demonstrate the surface-structure coupling geometry. }\label{fig:robot_skin}
%  \label{fig:flow_chart}
%\end{figure*}

%It has been extended to handle characters with more complex articulated skeletons by focusing on a time window in the %animation~\cite{Cohen:1992:ISC}.

The seminal work by Witkin and Kass~\cite{Witkin:1988} generated motion trajectories by optimizing physical constraints and animator controls at key frames, a well-known space-time constraints framework for animation. With proper motion data, the space-time optimization produces realistic articulated motions for bipedal or multilegged characters through different physical properties~\cite{Albro:845348,Cohen:1992:ISC,Crawford:M98/53,Fang:2003:ESP:882262.882286,Safonova:2004:SPR:1015706.1015754,Sulejmanpasic:2005:APB:1037957.1037966,Wampler:2009:OGF:1531326.1531366,Wei2011,Wampler:2014:GLS}. It can be used to transform motion capture data into physically plausible motions~\cite{Popovic:1999:PBM:311535.311536}.
%Integrating motion capture data can ease the user efforts to control the spacetime optimization, and

The locomotion controller aims to compute joint torques or control forces to drive the locomotion behaviors of articulated figures. The joint torques are usually calculated via the proportional and derivative (PD) controller such that the rigid skeleton of a character follows designated joint angle trajectories~\cite{Raibert:1991:ADL:127719.122755,Hodgins:1995:AHA:218380.218414,Coros:2011:LSS}. Balance control strategies, such as the swing foot placement or zero moment point, are essential to generating stable locomotions. ~\cite{Raibert:1991:ADL:127719.122755,Hodgins:1995:AHA:218380.218414,Mirua1994,Hodgins:1997:ASB:258734.258822,Kuo1999Stabilization,Kudoh2006,Niwa2009,Li2011}. Continuous adaptation of the target joint trajectory for balancing a walking human was developed in~\cite{Yin:2007:SSB:1276377.1276509}. Controllers that produce highly dynamic skills for human animation were suggested in~\cite{Liu:2010:SCM:1778765.1778865,Ha:2012:FLM:2366145.2366174,Liu:2012:TRC:2366145.2366173}. The joint torques can also be computed via optimal control to approximate the motion capture data or motion data from kinematic simulators~\cite{daSilva:2008:ISS:1360612.1360681,Muico:2009:CNC}.

Our work is inspired by studies on how to drive the soft skin deformation with the underyling rigid skeletons or pseudo muscle force~\cite{Jain:2011:CPC:2070781.2024197,Kim:2011:FSS,Tan:2012:SBL:2185520.2185522}.  Two-way coupling of rigid bodies and elastic bodies was considered in~\cite{Shinar:2008:TCR:1632592.1632607}. Fast simulation and control of soft robots of various configurations and actuations has also been studied using finite element method and the reduced formulation of compliance matrix~\cite{Duriez2013,Frederick2015,Christian2017,Coevoet2017,Bieze2018}.

%\noindent\textbf{Spacetime elastic body motion generation}
\vspace{5 pt}
\noindent\textbf{Elastic body simulation} focuses on the formulation of an elastic deformation energy and the proper handling of contact constraints to simulate realistic deformations of soft bodies~\cite{Terzopoulos:1987:EDM:37402.37427,Irving200666,hirota2000simulation,muller2002stable}. A comprehensive survey of physics-based elastic deformation models can be found in~\cite{nealen2006physically}.

Space-time optimization techniques can also be applied to control the motion of elastic bodies that are represented by volumetric meshes. To reduce the number of variables used to control the vertex positions in the optimization, model reduction techniques are frequently used~\cite{Barbic:2009,Barbic:2012:IED,Li:2014:SEE:2601097.2601217}. Barbi\v{c} et al.~\cite{Barbic:2009} imposed the equation of motion constraint in elastic body deformation, using the discrete adjoint method to compute the gradients of control forces. Pan et al.~\cite{Pan:2018:AAR} integrated the contact forces as additional variables to handle environment interactions and solved the space-time objective with alternating optimization, but did not handle the two-coupling problem we want to solve.

%Given the current control variable values, the adjoint method has two steps: a forward integration step to obtain the deformation and updated constraints and a backward pass to compute the gradient. The frame-by-frame optimization step in our algorithm can be viewed as a forward step in adjoin method to obtain the deformation of the skin mesh. Instead of a backward pass to compute gradient, we perform space-time optimization with simplified skin deformation information to obtain the motion.}

%Our work is also related to elastic simulation for character skinning in order to add secondary deformation effects to the animation. The skin deformation energy can be derived from the rigging parameters that control the shape deformation~\cite{Capell:2002:ISD:566654.566622,Capell:2005:PBR:1073368.1073412,Hahn:2012:RP:2185520.2185568,Hahn:2013:ESS:2485895.2485918}. McAdams et al.~\shortcite{McAdams:2011:EEC:2010324.1964932} developed an efficient elasticity model and a multigrid solver to obtain realistic skinning deformations for large-scale meshes.

%\input{overview}
%The coordination of

\section{Overview}
Given an input surface mesh of a quad-robot, we first design mechanical skeleton and skin mesh (\secref{sec:robot_design}) for the robot, and then use the proposed two-step alternating algorithm to optimize for a physically plausible motion plan (\secref{sec:locomotion_trajectory}). The overall system flowchart is illustrated in \figref{fig:teaser}.

During the alternating optimization, the first space-time optimization step outputs joint angle trajectories for the design according to the user-specified end-effector and COM trajectories (\secref{sec:spacetime_optimization}). This step is made possible by only considering the approximated skin deformation. The second frame-by-frame optimization step improves the physical plausibility of the joint angle trajectories with full simulation and various physical constraints(\secref{subsec:per-frame_optimization}), such as physical torque limits for the selected motors in the design. Two-way coupled multibody-elastic dynamics (\secref{sec:two-way}) are adopted in this step for the full simulation.

Finally, the designed robot is fabricated by fast prototyping methods for a physical validation (\secref{sec:robot_fabrication}), and stepper motors are mounted to drive the skeleton and the attached soft skin to realize the motion plan.

\section{Two-way Coupled Multibody-Elastic Dynamics}\label{sec:two-way}
%\todo{make sure when COM/COP first appears in the paper}
A core ingredient of our system is the dynamic simulation of the robot using a self-actuated rigid skeleton with a soft skin attached. We are inspired by existing coupled simulation systems in graphics~\cite{capell2002interactive,lee2009comprehensive,liu2013simulation} and exploit the Lagrange multipliers method to enforce the two-way coupling between the skeleton and the soft skin, which can be naturally integrated into the subsequent locomotion optimization.
%\bernd{the following sentence sounds a bit negative. Maybe instead saying: We build on successful work in graphics coupling the subsystems ..., and exploit the ...} Although coupling the subsystems of (multibody) rigid body dynamics and deformable simulation is not new in graphics ~\cite{capell2002interactive,lee2009comprehensive,liu2013simulation}, we exploit the Lagrange multiplier method to enforce the two-way  coupling between the skeleton and the soft skin, which can be naturally integrated into the subsequent locomotion optimization.
The Lagrange multipliers are used to guarantee that the skeleton and skin are attached to each other at prescribed locations, and we solve for all the unknown DOFs from both subsystems simultaneously.

%\begin{figure}[ht!]
%  \centering
%  \includegraphics[width=\columnwidth]{figs/glue/glue.pdf}
%  \caption{The two-way coupling between the rigid skeleton and the soft skin is enforced at prescribed glue vertices (red points). More precisely, as to be discussed in Sec.~\ref{sec:fabrication}, it is between the rubbery skin and the bracing unit. }
%  \label{fig:glue}
%\end{figure}

\subsection{Two-Way Skeleton-Skin Coupling}

We use Lagrangian mechanics for both the rigid skeleton and soft skin and obtain a symmetric formulation for these two subsystems. To ease the formulation of motion contraints, we use the generalized coordinates $\mathbf{q}$ to parameterize the entire skeleton, where $\mathbf{q} = \{c_x, c_y,c_z,q_1, ..., q_m\}$. The vector $\mathbf{q}$ is composed of the Cartesian coordinates of the center of mass (COM) at the root link $\{c_x,c_y,c_z\}$ and the three Euler angles at each joint $\{q_1,...,q_m\}$. We opt to model the soft skin using the neo-Hookean material model~\cite{ogden1972large} because it has been demonstrated to be well suited for robotic skins made out of silicone~\cite{Pozzi2018} and can handle large local deformations induced by joint rotation observed in our examples. If necessary, however, our approach should be easily extensible to more sophisticated material models such as Mooney-Rivlin and Ogden. The DOFs representing the vertex displacements of the tetrahedral skin mesh are denoted by the vector $\mathbf{u}$. Though numerical discretization, the  equations of the forward simulation of rigid skeleton and soft skin at each time step can be written into a linear system $\mathbf{A}x = \mathbf{b}$, where $\mathbf{A}$ is the system matrix and $\mathbf{b}$ is a vector of the sum of constant terms and external forces. The detailed derivation of two linear systems, $\mathbf{A}_r$, $\mathbf{b}_r$ for rigid skeleton and $\mathbf{A}_d$, $\mathbf{b}_d$ for soft skin, are elaborated in Appendix A.

%\Bernd{and has been demonstrated to be well suited for robotic skins made out of silicone~\cite{} and for the range of deformations.} \weiwei{I can not find a good reference. The related one I found via Google is froma a conference "International conference on systems and networks", not a good choice.}

The two subsystems are coupled by attaching the soft skin to the underlying skeleton at prescribed \emph{glue vertices}. Mathematically, this straightforward treatment leads to a set of nonlinear position constraints:
\begin{equation}\label{eq:pos_constraint}
\mathcal{C}(\mathbf{q},\mathbf{u})=\mathscr{R}(\mathbf{q})\mathbf{r}+\mathbf{t}-\mathbf{x}_c=0.
\end{equation}
The matrix $\mathscr{R}$ converts the positions on the rigid links where the skin is attached, $\mathbf{r}$, from local to world coordinates. This operation can be easily expressed as a rotation chain from the links all the way back to the root. Meanwhile, $\mathbf{t}$ concatenates the root translations of the rigid links, and $\mathbf{x}_c$ denotes the positions of the glue vertices on the skin mesh $\mathbf{x}_c = \mathbf{S}_c \mathbf{u}$, where $\mathbf{S}_c$ is a selection matrix.

Using the Lagrange multipliers method, we obtain the coupled multibody-elastic system:
%\begin{equation}
%\mathbf{A}=
%\left[
%\begin{array}{cc}
%\mathtt{diag}(\mathbf{A}_r, \mathbf{A}_d) & \nabla\mathcal{C}^\top\\
%\nabla\mathcal{C} & \mathbf{0}
%\end{array}
%\right],
%\;
\begin{equation}
\left[
\begin{array}{ccc}
\mathbf{A}_r  & \mathbf{0} & \nabla_q\mathcal{C}^\top\\
\mathbf{0}    & \mathbf{A}_d & \nabla_u\mathcal{C}^\top\\
\nabla_q\mathcal{C} & \nabla_u\mathcal{C} & \mathbf{0}
\end{array}
\right]
\left[
\begin{array}{c}
\Delta\mathbf{q}\\
\Delta\mathbf{u}\\
\pmb\lambda
\end{array}
\right]
=
\left[
\begin{array}{c}
\mathbf{b}_r\\
\mathbf{b}_d\\
\mathbf{0}
\end{array}
\right].
\label{eq:system_equation}
\end{equation}
%\todo{span a bit}
The coupling constraint is linearized via $\nabla\mathcal{C}$. In each time step, we solve for the changes of the system DOFs in Eq.~\ref{eq:system_equation}. Thus, we have $\mathbf{q}^{i} = \mathbf{q}^{i-1} + \Delta\mathbf{q}$ and $\mathbf{u}^{i} = \mathbf{u}^{i-1} + \Delta\mathbf{u}$, where the superscript $[\cdot]^{i}$ indicates the frame index.

%We linearize two nonlinear subsystems at each Newton iteration and couple them via $\nabla\mathcal{C}$.

\subsection{Collision and Contact Handling}
\label{subsec:collision_and_contact_handling}
There are two types of collisions/contacts we need to take care of in our simulation. The first is the collision between the robot's feet and the ground surface, which provides necessary support and friction forces to the robot. As will be detailed in Sec.~\ref{subsec:per-frame_optimization}, we resolve them using linear complementary constraints (LCP) to guarantee the physical feasibility of the locomotion.

The second is the self-collision of the soft skin; rotating joints tend to compress the inward skin and induce self-collisions. Moreover, skin-skeleton collisions\footnote{As discussed in Sec.~\ref{sec:fabrication},  skin-skeleton collisions can occur between the skin and the bracing unit, not the mechanical skeleton.} could also occur when the skeleton is being articulated. Because such collisions typically take place under low relative velocities, we handle them using the explicit penalty force method~\cite{Moore:1988:CDR, Bridson:2002:RTC}.

\section{Motion Plan Optimization}
\label{sec:locomotion_trajectory}
As the system input, the trajectories of the COM, end effectors, and the footfall pattern are provided by the user. Our system generates a dynamically feasible motion plan for the robot that resembles as much as possible the one prescribed by the user. A motion plan, defined as $\mathscr{P}=\{ \mathbf{q}^i, i=1...N; \Delta t \}$, includes the skeleton configuration $\mathbf{q}^i$ of the $i$-th frame for $i=1$ to $N$, and the time step size $\Delta t$.

\begin{figure}[t]
  \centering
  \includegraphics[width=\columnwidth]{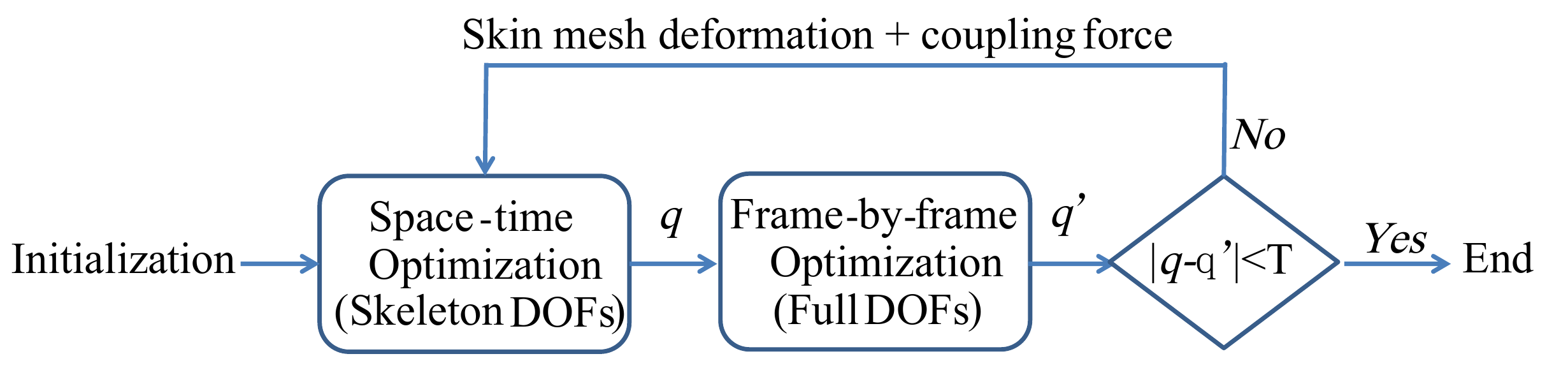}
  \caption{Alternating algorithm Flowchart.}
  \label{fig:algorithm_flowchart}
%\vspace{-16pt}
\end{figure}

%It is technically challenging to optimize all the frames in $\mathscr{P}$ entirely, as in~\cite{Megaro:2015:IDR}, due to the large number of DOFs in the multibody-elastic system and various physical constraints. For example, it is costly to minimize the required driving torques at joints in this optimization, since the torques should be computed through the inverse dynamics of the two-way coupled system.
Previous works, e.g. [Megaro et al 15], solved the motion design problem by simultaneously finding the optimal skeleton configuration for all the frames in $\mathscr{P}$. However, this problem becomes much more challenging in the case of a multibody-elastic system, due to the large number of DoFs and associated physical constraints. Thus, the global approach with full DOFs is infeasible in our case. In this section, we elaborate the details of our two-step alternating motion plan optimization algorithm which uses approximated skin deformation to significantly improve the efficiency of the global space-time optimization. Specifically, the influence of the skin deformation on the skeleton is approximated as the coupling forces at glue vertices and the influenced COM positions of each link. The convergence of the algorithm is tested considering the difference between the space-time optimized one and the optimized one after the frame-by-frame optimization step. \figref{fig:algorithm_flowchart} illustrates the algorithm flowchart.

%The first space-time optimization step takes care of the DOFs of the rigid skeleton, while considering the coupling forces at glue vertices and the influenced COM positions due to the skin deformation. The second frame-by-frame optimization step tries to follow the motion plan from the space-time optimization while full dynamics of two-way coupled system are simulated with various physical constraints. The second step provides the updated skin deformation to the first step and improves the physical plausibility of the optimized joint trajectories.

\subsection{Space-time Optimization With Skin Deformation Approximation}
\label{sec:spacetime_optimization}

%subject to: \hsapce{5pt} \mathcal{\phi}_e = 0, \mathcal{\phi}_i \leq 0,
%The variables in the space-time optimization that are collected at each frame $i$ are the set of generalized coordinates of the multi-body rigid skeleton $\mathbf{q}^i$, the contact forces $\mathscr{F}^i_{c,j}$ and the contact torques $\pmb\tau^i_c,j$ exerted on $j-$th end effector in contact with the ground.
%The optimization objective function has six terms to balance between the user-specified end effector and COM trajectories and the smoothness of the optimized motion:

Similar to the formulation in~\cite{Wampler:2009:OGF:1531326.1531366}, for each $i$-th frame, we consider the set of generalized coordinates of the rigid skeleton, $\mathbf{q}^i$, together with the contact forces $\mathscr{F}^i_{c,j}$ and torques $\pmb\tau^i_c,j$ that are exerted on the $j$-th end effector in contact with the ground. The influence of the skin deformation to the skeleton at $i$-th frame is simplified to be the coupling forces $\pmb{\lambda}^i$ at glue vertices and the skin deformation $\mathbf{u}^i$. They are obtained from the previous frame-by-frame optimization and not optimized in this step. The displacements in $\mathbf{u}^i$ are represented into the local coordinate frames of links at each frame in order to compute the influence of the skin deformation to the COM of the robot.

Given these quantities, the optimization objective is defined as the weighted sum of six terms that balance the user-specified end effector and COM trajectories and the smoothness of the optimized motion:
\begin{equation} \label{eq:space_time_optimization}
E_A = \min\sum_i\left(\alpha_tE_{\pmb\tau}^i+\alpha_sE_S^i+\alpha_cE_{COM}^i+\alpha_eE_{EE}^i+\alpha_fE_F^i+\alpha_oE_O^i\right).\
\end{equation}
The first term $E_{\pmb\tau}^i$ is standard in space-time optimization to minimize the torques $\pmb{\tau}^i$ exerted at the joints:
\begin{equation}
\nonumber
\displaystyle E_{\pmb\tau}^i = {\frac{1}{m}}^2\left\|\pmb{\tau}^i(\mathbf{q}^i, \pmb{\lambda}^i, \mathscr{F}^i_c, \pmb\tau^i_c)\right\|^2.
\end{equation}
The torques $\pmb{\tau}^i$ are computed using inverse dynamics, and the coupling forces $\pmb{\lambda}^i$ are integrated as the external forces exerted by the elastic skin at the coupling points.

The second term $E_{S}^i$ encourages the smoothness of the optimized motion, which is:
\begin{equation}
\nonumber
E_{S}^i = \left\|\mathbf{q}^{i+1} - 2\mathbf{q}^{i}+\mathbf{q}^{i-1}\right\|^2.
\end{equation}

\begin{wrapfigure}{r}{0.3\linewidth}
\vspace{-2 pt}
%\begin{center}
\includegraphics[width =\linewidth]{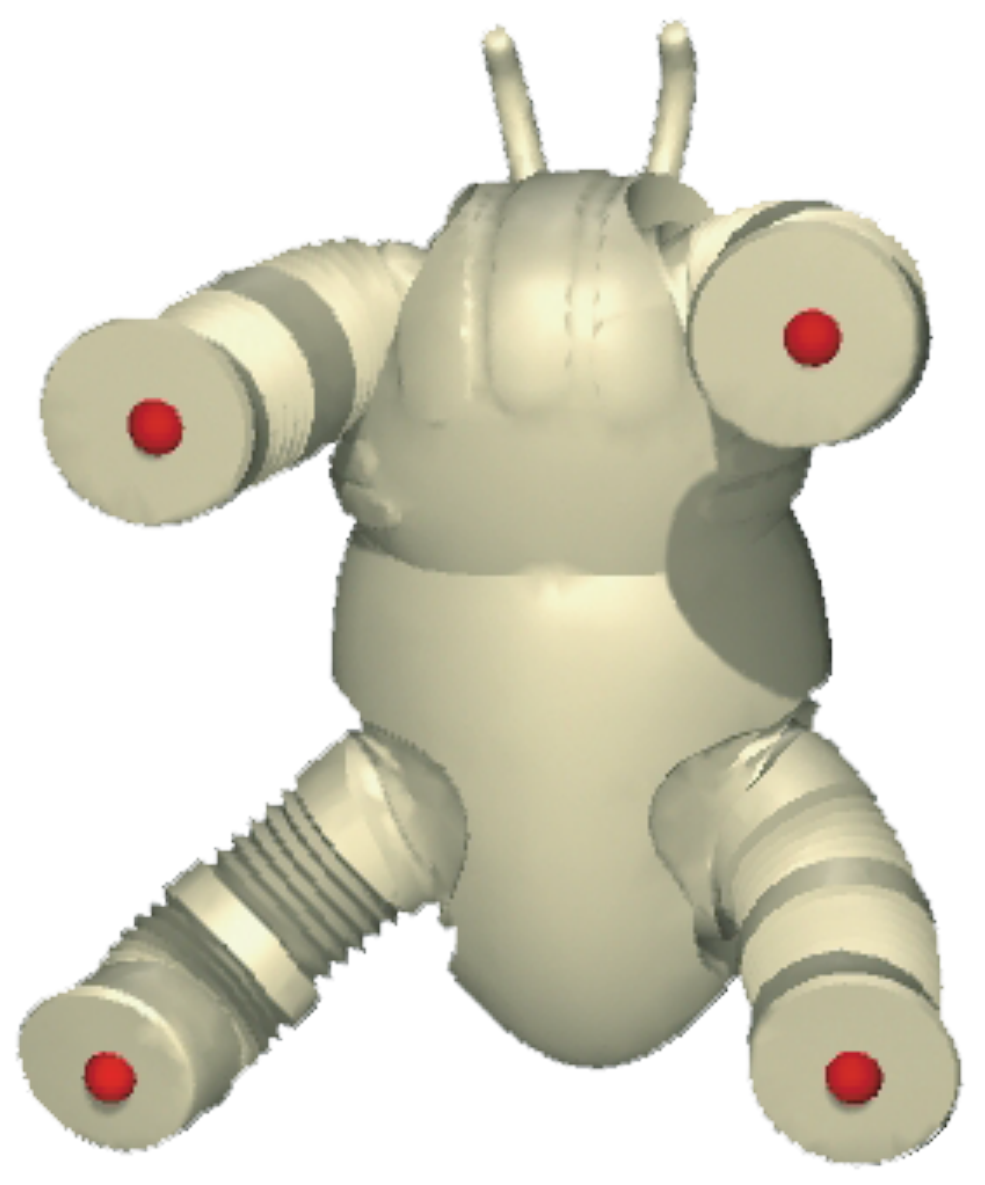}
\caption{Foot contacts.}
\label{fig:contact_vertices}
\vspace{-5 pt}
\end{wrapfigure}

The two terms, $E_{EE}^i$ and $E_{COM}^i$, enforce the optimized motion to follow the user-specified end-effector and COM trajectories respectively:
\begin{equation}
\nonumber
\begin{array}{ll}
\displaystyle E_{EE}^i = \left\|\phi_{EE}^i(\mathbf{q}^i,\mathbf{u}^i) - \mathbf{e}^i\right\|^2,&
\displaystyle E_{COM}^i = \left\|\phi^i_{COM}(\mathbf{q}^i,\mathbf{u}^i) - \mathbf{g}_i\right\|^2, \vspace{3 pt} \vspace{3 pt}\\
\end{array}
\end{equation}
where the functions $\phi_{EE}^i(\mathbf{q}^i,\mathbf{u}^i)$ and $\phi^i_{COM}(\mathbf{q}^i,\mathbf{u}^i)$ define how to compute the end effector and COM positions, given the generalized coordinates of the skeleton and the deformation of the skin mesh. For each end effector, we select one vertex closest to the end effector of the rigid skeleton and represent this vertex into the local coordinate system of the end effector to compute $\phi_{EE}$ (see \figref{fig:contact_vertices}).

The term $E_F^i$ penalizes the deviation from the motion $\{\tilde{\mathbf{q}}^i, i=1,..,N\}$ generated in the previous frame-by-frame optimization or the initialized motion plan at the first iteration:
\begin{equation}
\nonumber
E_F^i = \left\|\mathbf{q}^i - \tilde{\mathbf{q}}^i\right\|^2.
\end{equation}
After the frame-by-frame optimization step, the weight of the $E_F$ term to follow the motion plan of the previous iteration in the space-time optimization is increased by 10\%, which means the algorithm leans toward  physical plausibility.

The last term $E_O$ enforces that the end-effectors in the contact are flat:
\begin{equation}
\nonumber
\displaystyle E_{O}^i = \left\|\psi^i_{EE}(\mathbf{q}^i) - \hat{\mathbf{n}}\right\|^2,
\end{equation}
where $\psi^i_{EE}(\mathbf{q}^i)$ is a function to compute the orientation of the end effector and $\hat{\mathbf{n}}$ is set to be $(0,1,0)$, the normal of the support plane. The formulation of this term is motivated by the fact our skinned robot is soft and hence a contact area appears whenever feet contact the ground. We try to maximize the contact area at the moment of contact because it is important to achieve a stable motion. Similarly, when the foot is about to hover, we want to clear as many contact vertices as possible. Therefore, when an end effector is close to these important moments, we add $E_{O}$ to the objective function, which try to make the end effector face the upright direction of the ground.

%where:
%\begin{equation}\label{eq:objective_terms_expanded}
%\begin{array}{ll}
%\displaystyle E_{\pmb\tau}^i = {\frac{1}{m}}^2\left\|\pmb{\tau}^i(\mathbf{q}^i, %\pmb{\lambda}^i, \mathscr{F}^i_c, \pmb\tau^i_c)\right\|^2,&
%\displaystyle E_{S}^i = \left\|\ddot{\mathbf{q}}^i\right\|^2, \vspace{3 pt} \vspace{3 pt}\\
%\displaystyle E_{EE}^i = \left\|\phi_{EE}^i(\mathbf{q}^i,\mathbf{u}^i) - \mathbf{e}^i\right%\|^2,&
%\displaystyle E_{COM}^i = \left\|\phi^i_{COM}(\mathbf{q}^i,\mathbf{u}^i) - \mathbf{g}_i%\right\|^2, \vspace{3 pt} \vspace{3 pt}\\
%\displaystyle E_F^i = \left\|\mathbf{q}^i - \tilde{\mathbf{q}}^i\right\|^2, \vspace{3 pt}&
%\displaystyle E_{O}^i = \left\|\psi^i_{EE}(\mathbf{q}^i) - \hat{\mathbf{n}}\right\|^2, %\vspace{3 pt}\\
%\end{array}
%\end{equation}
%\bernd{should we add the constraints to/after equation (3), to show the optimization problem? just that people are not surprised about the constraints later on...}

We also impose hard kinematic and dynamic constraints on the optimized variables to obtain a stable motion plan. The kinematic constraints include the contact constraint, the center of pressure (COP) constraint and the optional periodic constraint, while the dynamic constraints include the momentum and the friction force constraints as in ~\cite{Wampler:2009:OGF:1531326.1531366}.

% \bernd{This figure could be improved. Suggestions: Replace Initialization with the actual input, and end with the output. Also: Could the font size be increased a little bit?}} \weiwei{I increased the font size.}

\noindent\textbf{Contact constraint}: In the optimization, we need to enforce the footfall pattern that is specified by the user to encode when the foot should leave or touch the ground. This constraint can be written into:
\begin{equation}\label{eq:foot_contact_constraint}
\begin{array}{ll}
\displaystyle c^i_j{\phi_{EE,j}^i(\mathbf{q}^i,\mathbf{u}^i)}_y=0,& \forall i,j,\\
\displaystyle c^{i-1}_jc^i_j(\phi_{EE,j}^{i}(\mathbf{q}^i,\mathbf{u}^i) - \phi_{EE,j}^{i-1}(\mathbf{q}^i,\mathbf{u}^i)) = 0,& \forall i,j,\\
\end{array}
\end{equation}
where $c^i_j$ is a binary variable set to 1 if the $j$-th end effector is in contact with the ground at the $i$-th frame. This variable can be directly derived from the footfall pattern. The $y$ coordinate of the end effector, denoted by ${\phi_{EE,j}^i(\mathbf{q}^i,\mathbf{u}^i)}_y$, should be $0$ since the ground is set to be $y=0$.

\noindent\textbf{COP constraint}: The COP should be inside the supporting polygon for a stable motion plan. This constraint can be written into:
\begin{equation} \label{eq:COP_Constraint}
 \displaystyle\mathbf{P}\cdot\phi^i_{COP} \leq \mathbf{0}, \hspace{80pt} \forall i,
\end{equation}
where the funtion $\phi^i_{COP}$ computes the COP position using the same method as in~\cite{Megaro:2015:IDR}. The rows in $\mathbf{P}$ represent edges of the supporting polygon. Since the space-time optimization requires the foot to be flat on the supporting plane, the supporting polygon is formed by the convex hull of the vertices that represent the sole meshes of the end-effectors in contact. However, the contact forces and torques at a sole are simplified to be exerted on a single point of the end effector to ease the optimization. \figref{fig:contact_vertices} illustrates the contact points (red balls) selected as the positions of the end-effect for the beetle-like robot.

%and the function $\mathtt{proj}(\cdot)$ projects the position of COP onto the ground surface by simply discarding the $y$ component of the computed COP position.

\noindent\textbf{Periodic constraint}: When the user desires a periodic motion, the joint angles are expected to be the same in the first and the last frame of the optimized motion plan:
\begin{equation}
J(q^1) = J(q^N),
\end{equation}
where $J$ extracts the joint angles from the generalized coordinates.

\noindent\textbf{Momentum constraint}: The change of the linear and angular momentum of the robot should be determined by the external contact forces and torques, which can be formulated into the following equations:
\begin{equation}
\begin{array}{ll}
\displaystyle \dot{\mathbf{R}}^i = mg + \sum_jc^i_j\mathscr{F}^i_{c,j}, & \forall i,j,\\
\displaystyle \dot{\mathbf{L}}^i = \sum_j c^i_j\left(\left(\mathbf{p}_{c,j}^i - \phi^i_{COM}(\mathbf{q}^i,\mathbf{u}^i)\right)\times \mathscr{F}^i_{c,j}+ \pmb\tau_{c,j}^i\right), & \forall i,j,\\
\end{array}
\end{equation}
where $\mathbf{R}^i$ and $\mathbf{L}^i$ are, respectively, the total linear and angular momentum of the robot at $i$-th frame, and $\mathbf{p}_{c,j}^i$ gives the contact position of the $j$-th end effector.

\noindent\textbf{Friction force constraint}: The contact force should be inside the friction cone to satisfy the Coulomb model of friction. Specifically, we have:
\begin{equation}
\begin{array}{ll}
\displaystyle\frac{1}{m}(\mu{\mathscr{F}^i_{c,j}}_\perp - \|{\mathscr{F}^i_{c,j}}_{\|}\|) \geq 0, & \forall c^i_j=1,\\
  \displaystyle\frac{1}{m}^2(\|{\mathscr{F}^i_{c,j}}\|^2 - \|\frac{{\pmb\tau^i_j}_{\|}}{\nu_b}\|^2 - \frac{{\pmb\tau^i_j}_{\perp}}{\nu_t}^2) \geq 0, & \forall c^i_j=1, \\
\end{array}
\end{equation}
where ${\mathscr{F}^i_{c,j}}_\perp$ and ${\mathscr{F}^i_{c,j}}_\|$ represent the components of the contact force perpendicular and parallel to the ground respectively. The first equation requires the contact force to be inside a friction cone. The second equation relates the contact force and the contact torque, where $\nu_b$ and $\nu_t$ are set to be the radius of the circumcircle of the sole mesh.

\noindent\textbf{Optimization}: With the defined objective function and constraints, the space-time optimization step can be written into:
\begin{equation}
\begin{array}{ll}
\min_{\mathbf{q}^i,\mathscr{F}^i_{c,j}, \pmb\tau^i_c,j, i=1,..N} & E_A\\
\text{st.}: \hspace{5pt} \mathbf{\phi}_e = 0, \mathbf{\phi}_g \leq 0 &
\end{array}
\end{equation}
where $\mathbf{\phi}_e$ and $\mathbf{\phi}_g$ represent the equality and inequality constraints respectively. We solve this sequential quadratic programming problem using the Gauss-Newton algorithm. The weights in the objective function are specified as follows: $\alpha_t = 1e-2, \alpha_s = 0.5, \alpha_c = \alpha_e = 1$, $\alpha_f = 1$ and $\alpha_o = 10$.

%$\sum_i\|\pmb{\tau}^i(\mathbf{\pmb\lambda}^i, \mathbf{q}^i)\|^2+w_2\|\ddot{\mathbf{q}}^i\|^2+w_3\|\mathbf{q}^i - \tilde{\mathbf{q}}^i\|^2+\left\|\phi^i_{COM} - \mathbf{g}_i\right\|^2+ \left\|\phi_{EE}^i - \mathbf{e}^i\right\|^2+\left\|\psi^i_{EE} - \hat{\mathbf{n}}\right\|^2

\subsection{Frame-by-frame Optimization with Full Dynamics}
\label{subsec:per-frame_optimization}
In the frame-by-frame optimization, we drop the simplification of the previous step and consider the full dynamics formulated in Eq.~\eqref{eq:system_equation} in a similar way as in \cite{Tan:2012:SBL:2185520.2185522}. This allows us to further improve the physical plausibility of the motion. In practice, this requires the solution of a difficult quadratic programming problem with complementarity constraints (QPCC) to handle contact and friction. Different to the soft body only formulation in \cite{Tan:2012:SBL:2185520.2185522}, the coupling constraints between multi-body skeletion and elastic skin introduce a large number of Lagrange multiplier variables and largely increase the complexity of the solver. To speed up the solution, we follow the condensation technique widely used in physical simulation~\cite{Nielsen1996,Teng:2015:SCF}. Specifically, we select the driving torques at joints and foot contact forces as optimization variables and lump the DOFs of mesh vertices and rigid skeletons to these variables through the condensation of the system matrix in Eq.~\eqref{eq:system_equation}. This choice facilitates the formulation of the physical torque limit constraints of motors as well. In this section, we describe the details of the matrix condensation and the per-frame optimization problem.

\noindent\textbf{System matrix condensation}: With the coupled multibody-elastic system defined in Eq.\eqref{eq:system_equation}, the nonlinear relation between the simulated DOFs $\Delta\mathbf{u}$  and $\Delta\mathbf{q}$ and the vector $\mathbf{b}_r$ and $\mathbf{b}_d$ can be revealed by eliminating the unknown Lagrange multipliers $\pmb{\lambda}$ in the matrix condensation (see Appendix.B for the detailed derivation):
\begin{equation}\label{eq:condensation_bu}
\begin{array}{l}
\Delta\mathbf{u}=\mathbf{A}_d^{-1}\mathbf{b}_d-\mathbf{A}_d^{-1}\nabla_u\mathcal{C}^\top\mathbf{A}_C^{-1}\left(\nabla_q\mathcal{C}\mathbf{A}_r^{-1}\mathbf{b}_r+\nabla_u\mathcal{C}\mathbf{A}_d^{-1}\mathbf{b}_d\right),\\
\Delta\mathbf{q}=\mathbf{A}_r^{-1}\mathbf{b}_r-\mathbf{A}_r^{-1}\nabla_q\mathcal{C}^\top\mathbf{A}_C^{-1}\left(\nabla_q\mathcal{C}\mathbf{A}_r^{-1}\mathbf{b}_r+\nabla_u\mathcal{C}\mathbf{A}_d^{-1}\mathbf{b}_d\right),
\end{array}
\end{equation}
where $\mathbf{A}_C=\nabla_q\mathcal{C}\mathbf{A}_r^{-1}\nabla_q\mathcal{C}^\top+\nabla_u\mathcal{C}\mathbf{A}_d^{-1}\nabla_u\mathcal{C}^\top$.

According to the formulation of $\mathbf{b}_r$ and $\mathbf{b}_d$ in Appendix.A, the joint torques and the contact forces exerted on the soft skin are absorbed into the vector $\mathbf{g}_r$ and $\mathbf{g}_d$. Since the rest terms in $\mathbf{b}_r$ and $\mathbf{b}_d$ are constant, we can use $\phi_{\Delta\mathbf{q}}$ and $\phi_{\Delta\mathbf{u}}$ to represent Eq.~\eqref{eq:condensation_bu} more precisely:
\begin{equation}\label{eq:phi_q_phi_u}
\Delta\mathbf{q}=\phi_{\Delta\mathbf{q}}\big(\pmb{\tau},\mathscr{F}_{\perp},\mathscr{F}_{\|}\big),\quad\text{and}\quad\Delta\mathbf{u}=\phi_{\Delta\mathbf{u}}\big(\pmb{\tau},\mathscr{F}_{\perp},\mathscr{F}_{\|}\big).
\end{equation}
Here, $\mathscr{F}_{\perp}$ and $\mathscr{F}_{\|}$ are magnitudes of normal and tangent forces at all the contact vertices on the skin mesh, and $\pmb{\tau}$ represnets the join torques. To handle LCP constraints, the contact force at a contact vertex is modeled as $\mathscr{F}=\hat{\mathbf{n}}\mathscr{F}_{\perp}+\mathbf{D}\mathscr{F}_{\|}$ instead of the 3D vector representation in~\secref{sec:spacetime_optimization}, where $\hat{\mathbf{n}}\mathscr{F}_{\perp}$ represents the upright supporting force along the contact normal $\hat{\mathbf{n}}=[0,1,0]^\top$ and $\mathscr{F}_{\perp}\in\mathbb{R}$ is the force magnitude. $\mathbf{D}\in\mathbb{R}^{3\times4}$ is a matrix and its columns are the vectors that span the contact plane. In our system, we use four directions to form the friction cone~\cite{Anitescu1997} and $\mathscr{F}_{\|}\in\mathbb{R}^4$ is the tangent magnitude. The explicit penalty forces for resolving the collision between the skeleton and skin are pre-determined quantities and merged into the constant terms in $\mathbf{b}_r$ and $\mathbf{b}_d$.

Given $\mathbf{q}^{i-1}$ and $\mathbf{u}^{i-1}$, $\Delta\mathbf{q}^i$ and $\Delta\mathbf{u}^i$ determine the positions and orientations of COM, COP, and end effectors at the $i$\textsuperscript{th} frame. Thanks to Eq.~\eqref{eq:phi_q_phi_u}, these kinematics variables are now functions of $\pmb{\tau}$, $\mathscr{F}_{\perp}$, and $\mathscr{F}_{\|}$. Therefore, we can derive  the functions required in the computation of kinematic information, namely $\phi^i_{EE}, \psi^i_{EE}, \phi^i_{COM}$ and  $\phi^i_{COP}$, if external forces $\pmb{\tau}^i$, $\mathscr{F}_{\perp}^i$, and $\mathscr{F}_{\|}^i$ are given. Note that the purpose of these functions are explained in the space-time optimization (see Sec.~\ref{sec:spacetime_optimization}).

\begin{figure}[t]
  \centering
  \includegraphics[width=\columnwidth]{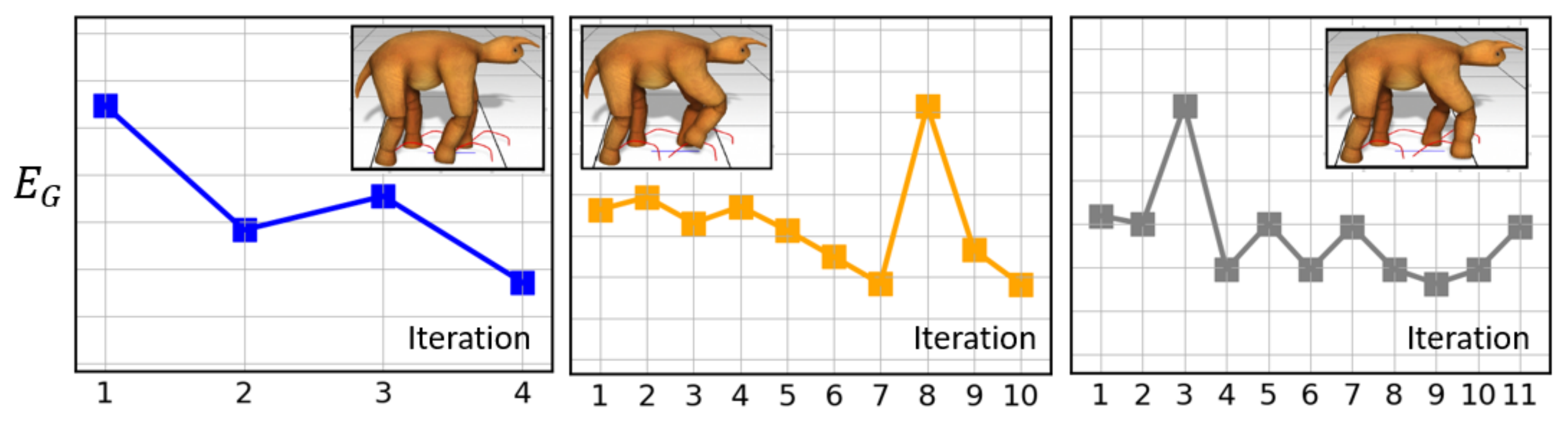}
  \caption{Our QPCC solver converges quickly in most cases. The left plot is the converging curve of a frame when the front left leg of the monster-like robot leaves the ground. The middle plot is the converging curve of a frame when this leg is in the air (i.e. other three feet are on the ground). The right plot is the converging curve of a frame when this leg hits ground again.}
\label{fig:qpcc}
\end{figure}

\noindent\textbf{Optimization}: We follow the control strategy used in~\cite{Tan:2012:SBL:2185520.2185522} to optimize the input motion plan on a frame-by-frame basis. It is used to make sure that the output joint angle trajectories of the space-time optimization step are physically feasible, which is achieved using the two-way coupled multibody-elastic dynamics as constraints. At each frame, it can be formulated as a quadratic programming problem with complementarity constraints.

%We further simplify the pivoting procedure between LCPs in order to accelerate the optimization.
%Specifically, we seek for joint torques and contact forces so that the corresponding $\Delta \mathbf{q}^i$ and $\Delta \mathbf{u}^i$ to step forward at frame $i$. The optimization problem at frame $i$ can be formulated into:
Specifically, we seek for joint torques ($\pmb{\tau}^i$) and contact forces ($\mathscr{F}_{\perp}^i$, $\mathscr{F}_{\|}^i$) such that the corresponding $\Delta \mathbf{q}^i=\phi^i_{\Delta\mathbf{q}}$ and $\Delta \mathbf{u}^i=\phi^i_{\Delta\mathbf{u}}$ satisfy necessary hard constraints and the resulting locomotion matches the input locomotion as much as possible. Mathematically, it is formulated as
\begin{equation}\label{eq:original_optimization}
\begin{array}{lll}
&\displaystyle\min_{\pmb{\tau}^i,\mathscr{F}_{\perp}^i,\mathscr{F}_{\|}^i,\pmb{\lambda}_{\|}^i} E_G\left(\pmb{\tau}^i,\mathscr{F}^i_{\perp},\mathscr{F}^i_{\|}\right) & \vspace{5 pt} \\
& \displaystyle\text{subject to:}  \vspace{5 pt}\\
& \displaystyle\|\pmb{\tau}^i_m\| < U_m, m = 1,...,M  \\
& \displaystyle\mathbf{P}\cdot\phi^i_{COP}(\pmb{\tau}^i,\mathscr{F}_{\perp}^i,\mathscr{F}_{\|}^i) \leq \mathbf{0}  \vspace{5 pt}\\
& \displaystyle
%\left{
\mathbf{0} \leq
\left[
\begin{array}{c}
\mathscr{F}^i_{\perp} \vspace{5 pt}\\
\mathscr{F}^i_{\|} \vspace{5 pt}\\
\lambda^i_{\|}
\end{array}
\right]
\perp
\left[
\begin{array}{c}
\displaystyle \hat{\mathbf{n}}^\top\frac{\Delta\mathbf{u}_c^i}{\Delta t} \\ \displaystyle \mathbf{D}^{\top}\frac{\Delta\mathbf{u}_c^i}{\Delta t} + \displaystyle \mathbf{1}\lambda_{\|}\\
\mu \mathscr{F}^i_{\perp} - \mathbf{1}^{\top}\mathscr{F}^i_{\|}
\end{array}
\right] \geq \mathbf{0}.
%\right}.
\end{array}
\end{equation}
%\begin{equation}
%\begin{aligned}
%\min_{\pmb{\tau},\mathbf{f}_{\perp},\mathbf{f}_{\parallel},\lambda_L} E_G\left(\pmb{\tau},\mathbf{f}_{\perp},\mathbf{f}_{\parallel}\right) \hspace{80pt}\\\
%\text{Subject to:}\hspace{120pt}\\
%\|\mathbf{\tau}_i\| < U_{\tau}, i = 1,..,N_j \hspace{40pt}\\
%\mathbf{B}\phi_{COP}^h \leq 0 \hspace{80pt}\\
%0 \leq \left[ \begin{array}{c}\mathbf{f}_{\perp} \\ \mathbf{f}_{\parallel} \\ \lambda_l\end{array} \right] \perp \left[ \begin{array}{c}\mathbf{N}^\top\frac{\Delta\mathbf{u}_c^i}{\Delta t} \\ \mathbf{D}^{\top}\frac{\Delta\mathbf{u}_c^i}{\Delta t} + \mathbf{E}\lambda_l \\ \mu \mathbf{f}_{\perp} - \mathbf{E}^{\top}\mathbf{f}_{\parallel} \end{array} \right] \geq 0\\
%\end{aligned}
%\label{eq:original_optimization}
%\end{equation}
In Eq.~\eqref{eq:original_optimization}, the first hard inequality constraint of $\|\pmb{\tau}^i_m\| < U_m$ is the motor constraint requiring for all the $M$ motors that the computed torque magnitude is within its physical limit $U_m$.
% While $\pmb{\tau}_m\in\mathbb{R}^3$ is a vector, in practice we can represent it with a single scalar as only uniaxial motors are used in our fabrication.
The second inequality constraint $\mathbf{P}\cdot\phi_{COP} \leq \mathbf{0}$ requires the position of the COP to be within the supporting polygon as in Sec.~\ref{sec:spacetime_optimization}. The last complementary constraint is enforced at each individual contact vertex. It characterizes the contact mechanism such that when normal force exists, the relative velocity between the ground and the contact vertex along the contact normal should be zero, etc. Here, $\Delta\mathbf{u}^i_c\in\mathbb{R}^3$ is the incremental displacement of a contact vertex. The auxiliary vector $\lambda_{\|}$ is related to the tangent velocity of a sliding contact; $\mu$ is the friction coefficient; and $\mathbf{1}$ is a vector of ones, that is, $\mathbf{1}=[1,1,1,1]^\top$.
%
%Since the contact points of the robot are from the soft skin mesh, the linear complementary constraints for contact handling are constructed with $\Delta\mathbf{u}_c^i$ and additional multiplier $\lambda_l$, where $\Delta\mathbf{u}_c^i$ indicates the collected DOFs at contact points.

%\displaystyle E_{COM} = \left\|\phi^i_{COM} - \bar{\mathbf{g}}_i%\right\|^2, \vspace{3 pt} \vspace{3 pt}&
%\displaystyle E_{EE} = \left\|\phi_{EE}^i - \mathbf{e}^i\right\|^2, \\

%$E_{COM}$ pushes the COM of the resulting pose toward the input COM trajectory ($\mathbf{g}^i$) given by the user. $E_{EE}$ tries to make the trajectories of end effectors approximate the one prescribed by the user ($\mathbf{e}^i$).

The objective function $E_G$ has four terms:
\begin{equation}\label{eq:objective_terms}
E_G= \alpha_S E_S + \alpha_F E_F + \alpha_{O} E_{O} + \alpha_{\pmb{\tau}}E_{\pmb{\tau}} +\alpha_C E_C,
\end{equation}
where:
\begin{equation}\label{eq:objective_terms_expanded}
\begin{array}{ll}
\displaystyle E_S = \left\|\phi^i_{\Delta\mathbf{q}} - \phi^{i-1}_{\Delta\mathbf{q}}\right\|^2, &
\displaystyle E_F = \left\|\mathbf{q}^{i} + \phi^i_{\Delta\mathbf{q}}  - \bar{\mathbf{q}}^{i+1}\right\|^2,\\
\displaystyle E_{O} = \left\|\psi^i_{EE} - \hat{\mathbf{n}}\right\|^2, \vspace{3 pt} &
\displaystyle E_{\pmb{\tau}} = \left\|\pmb{\tau}_i - \pmb{\tau}_{i-1}\right|^2\\
\displaystyle E_{C} = \left\|\mathscr{E}\left(\phi_{EE}^i - \phi_{EE}^{i-1}\right)\right\|^2.
\end{array}
\end{equation}

The first energy term $E_S$ is the smoothness penalty, which favors motions with consistent velocities. The term $E_F^i$ penalizes the deviation from the motion $\{\bar{\mathbf{q}}^i, i=1,..,N\}$ generated in the previous space-time optimization. $E_{O}$ is the same soft constraint on the orientation of an end effector as in Eq.~\eqref{eq:space_time_optimization}, which can maximize the contact area for a stable motion. $E_{\pmb{\tau}}$ is used to penalize the large variation of the control torques at joints between frames. The last term $E_{C}$ imposes a penalty to moving end effectors who are responsible for supporting feet. In other words, if a foot is in contact with the ground and supporting the body, we use $E_{C}$ to reduce the risk of its possible tangent sliding. Here, $\mathscr{E}$ is an elementary matrix that picks positions of supporting end effectors. The weighting constants for each of these penalty terms are as follows: $\alpha_{S}=1$, $\alpha_{F}=10$, $\alpha_{O}=2$, $\alpha_{\pmb{\tau}} = 0.5$ and $\alpha_{C}=10$.

\subsection{Solving the QPCC}
The key to solving the QPCC problem of Eq.~\eqref{eq:original_optimization} is to have a feasible configuration for all the contact vertices. Our strategy is similar to that of ~\cite{Tan:2012:SBL:2185520.2185522}: We flip complementary constraints when the inequality constraint reaches the boundary. Specifically, contact vertices fall into one of the three following categories:

\begin{itemize}[leftmargin=10pt]
\item \textbf{Contact breakage} means that the contact vertex will leave the ground plane in the next frame, and the complementary constraints should be lifted.
\item \textbf{Sliding} indicates that the contact vertex is moving within the ground plane. In this situation, the complementary constraint for its normal force $\mathscr{F}_{\perp}$ becomes:
\begin{equation}\label{eq:sliding_constraint_normal}
\mathscr{F}_{\perp} > 0,\quad\hat{\mathbf{n}}^\top \frac{\Delta\mathbf{u}_c^i}{\Delta t} = 0,
\end{equation}
and the complementary constraints for the tangent force $\mathscr{F}_\parallel$ are:
\begin{equation}\label{eq:sliding_constraint_tangent}
\begin{array}{ll}
\displaystyle\mathscr{F}_\parallel \geq \mathbf{0}, & \displaystyle\mathbf{D}^{\top}\frac{\Delta\mathbf{u}_c^i}{\Delta t} + \mathbf{1}\lambda_{\|} = \mathbf{0}; \vspace{3 pt}\\
\displaystyle\lambda_{\|} \geq 0, &
\displaystyle\mu \mathscr{F}_{\perp} - \mathbf{1}^{\top}\mathscr{F}_{\|} = 0.
\end{array}
\end{equation}
%where $\mu$ is the dynamic friction coefficient.
\item \textbf{Static friction} implies that the contact vertex is fixed on the contact plane. In this case, the complementary constraint for its normal force is the same as Eq.~\eqref{eq:sliding_constraint_normal}. The constraints for the tangent force are:
\begin{equation}\label{eq:static_constraint}
\begin{array}{cc}
\displaystyle\mathscr{F}_\parallel \geq \mathbf{0}, & \displaystyle\mathbf{D}^{\top}\frac{\Delta\mathbf{u}_c^i}{\Delta t} + \mathbf{1}\lambda_{\|} = \mathbf{0}; \vspace{3 pt}\\
\displaystyle\lambda_{\|} = 0, &
\displaystyle\mu \mathscr{F}_{\perp} - \mathbf{1}^{\top}\mathscr{F}_{\|} \geq 0.
\end{array}
\end{equation}
%where $\mu$ is the static friction coefficient.
The inequality constraint of $\mu \mathscr{F}_{\perp} - \mathbf{1}^{\top}\mathscr{F}_{\|} \geq 0$ specifies the friction cone constraint in the case of static friction.
d\end{itemize}

We begin solving Eq.~\eqref{eq:original_optimization} by assuming all the contact vertices are fixed, which simplifies the original problem to
\begin{equation}\label{eq:initial_optimization}
\begin{array}{lll}
&\displaystyle\min_{\pmb{\tau}^i,\pmb{\gamma}_c} E_G\left(\pmb{\tau}^i,\pmb{\gamma}_c\right) & \vspace{5 pt} \\
& \displaystyle\text{subject to:}  \vspace{5 pt}\\
& \displaystyle\|\pmb{\tau}^i_m\| < U_m, m = 1,...,M  \\
& \displaystyle\mathbf{P}\cdot\phi^i_{COP} \leq \mathbf{0}  \vspace{3 pt}\\
& \displaystyle\left[\phi^i_{\Delta\mathbf{u}}\right]_c=\mathbf{0},
\end{array}
\end{equation}
where $\left[\phi^i_{\Delta\mathbf{u}}\right]_c$ returns the incremental displacements of all the contact vertices. This assumption of fixing all the contact vertices is realized via the Lagrange multipliers method, and the resulting multipliers $\pmb{\gamma}_c$ correspond to the constraint forces at these vertices. Now, let $\pmb{\gamma}_c\in\mathbb{R}^3$ be the constraint force at one of the contact vertices. It can be decomposed along normal and tangent directions as:
\begin{equation}\label{eq:lambda_decomposition}
\gamma_{\perp} = \hat{\mathbf{n}}^\top\pmb{\gamma}_c,\quad\text{and}\quad
\pmb{\gamma}_{\|} = \big(\mathbf{I}-\hat{\mathbf{n}}\hat{\mathbf{n}}^\top\big)\pmb{\gamma}_c.
\end{equation}
We label all the contact vertices as contact breakage, static friction, or sliding by checking $\gamma_{\perp}$ and $\pmb{\gamma}_{\|}$. If $\gamma_{\perp}\leq 0$, which indicates a contact breakage, we remove the constraint at the vertex in the next iteration. If $\gamma_\perp > 0$, we further examine the magnitudes of $\mu\gamma_{\perp}$ and $\|\pmb{\gamma}_{\|}\|$. If $\mu\gamma_{\perp}>\|\pmb{\gamma}_{\|}\|$, the vertex falls into the static friction category, otherwise the vertex is considered sliding. After all the contact vertices are labelled, we can convert the complementary constraints into a set of equality or inequality constraints, as explained in Eqs.~\eqref{eq:sliding_constraint_normal}, \eqref{eq:sliding_constraint_tangent}, and \eqref{eq:static_constraint}, and re-solve the QP optimization. It is known that QPCC is NP-complete, and few contact vertices could make the optimization procedure computationally intractable. Therefore, we simplify this procedure by grouping vertices on the planar surface of the foot mesh into five patches similar to~\cite{Tan:2012:SBL:2185520.2185522}. In our experiments, we found that such initial vertex grouping often provide a good start for the QPCC solver. Typical converging curves are plotted in \figref{fig:qpcc}, and we stop the optimization after 10 iterations. We observe that the condensed QPCC solver is around 50x faster than the QPCC without condensation.

\subsection{Initialization}
Given the mechanical skeleton and the skin mesh of a robot, we first associate the mesh vertices to the links of the skeleton to obtain its skinning information. Hence, the mesh vertices can be deformed with the skeleton in the space-time optimization step, while the local coordinates of the vertices should be computed using their deformed positions and the local frame of the links at each frame. The initial motion plan are computed through the space-time optimization step without the trajectory following terms $E_F^i$ . In this step, the skin deformation is assumed to be static and each link has additional weights from its associated vertices. Afterwards, the initial skin mesh deformation is simulated by imposing the coupling constraints in the elastic simulation of the skin, and the initial coupling force is then obtained according to the deformation of the tetrahedra connected to the coupling points~\cite{Kim:2011:FSS}.

\section{Design and Fabrication}
\label{sec:fabrication}
\begin{figure}[t!]
  \centering
  \includegraphics[width=\linewidth]{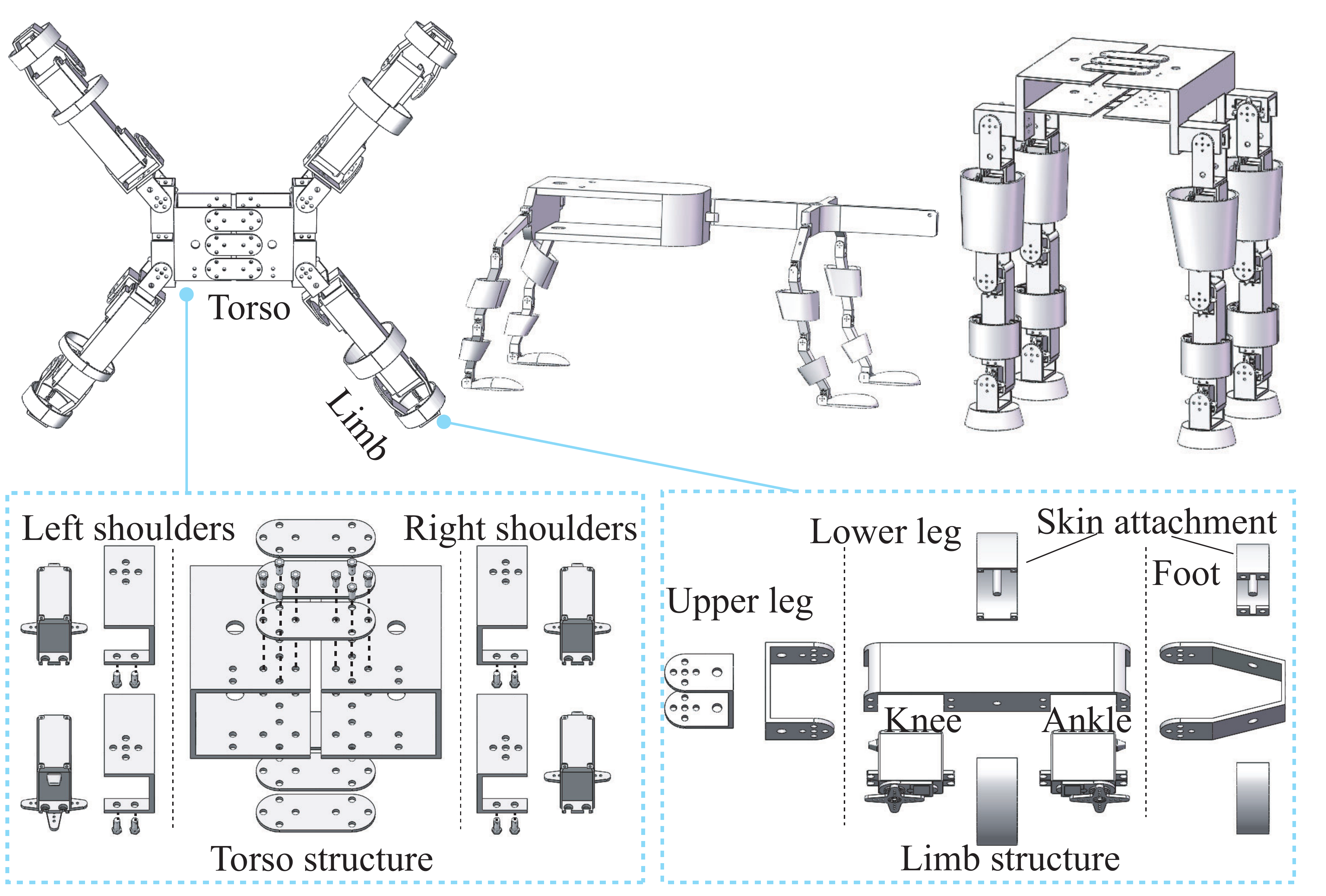}
  \caption{Three mechanical structure templates and the exploded views of the torso and limb structures of the first template.}
  \label{fig:templates}
\end{figure}

Designing and fabricating a quad-robot is a challenging task. We facilitate the design by using a set of mechanical skeleton templates, and narrow the gap between professional and regular users by creating several SolidWorks scripts. This allows even an inexperienced user to tweak high-level semantic parameters. \figref{fig:templates} shows three built-in mechanical skeleton templates provided in our system for quad-robots. Each template is built of modularized CAD parts to ease the fabrication cost. The first one is the design used in the beetle-like robot, which consists of a torso structure and four limb structures. Their exploded views are detailed in the figure as well.
%\noindent \textbf{Template mechanical structure}: Three quad-robot mechanical structure templates are illustrated in \figref{fig:templates}. The first template is a basic design used in our beetle-like robot, which consists of a torso structure and four limb structures. The other two templates vary with different initial poses, foot part and additional joints and links at the torso. The geometry of each template is designed in \texttt{SolidWorks}.
The torso structure has four shoulder joints that connect to its four limbs. A microcontroller board sending trigger signals to the motors is mounted inside of the torso. The limb structure includes linkage parts of an upper leg, a lower leg, and a foot. On each limb, two uniaxial motors are mounted to provide necessary rotational freedoms at the knee and the ankle. The other two templates vary in different initial poses and foot link geometries.

In the following, we describe the details of the design pipeline and the fabrication procedure respectively.

%If necessary, the user can edit the geometry of a modular part to create a customized skeleton design. We also developed several \texttt{SolidWorks} scripts, which allow an inexperienced user to tweak high-level semantic parameters of a CAD part conveniently. The design of the skin is primarily based on a user-specified 3D robot model. We fabricate the skin piece by piece, attach them to the mechanical skeleton at prescribed locations, and glue them together. To avoid large stretching forces induced by joint bending, our system also allows the user to add folding regions on the original skin surface, which increases surface area and relieves localized skin deformations. We fabricate the mechanical skeleton using 3D printing and the exterior soft skin using injection molding.

\subsection{Design and Editing of Mechanical Skeleton}
\label{sec:robot_design}
\begin{wrapfigure}{r}{0.3\linewidth}
\vspace{-5 pt}
%\hspace{-1.35cm}
%\begin{center}
\includegraphics[width =\linewidth]{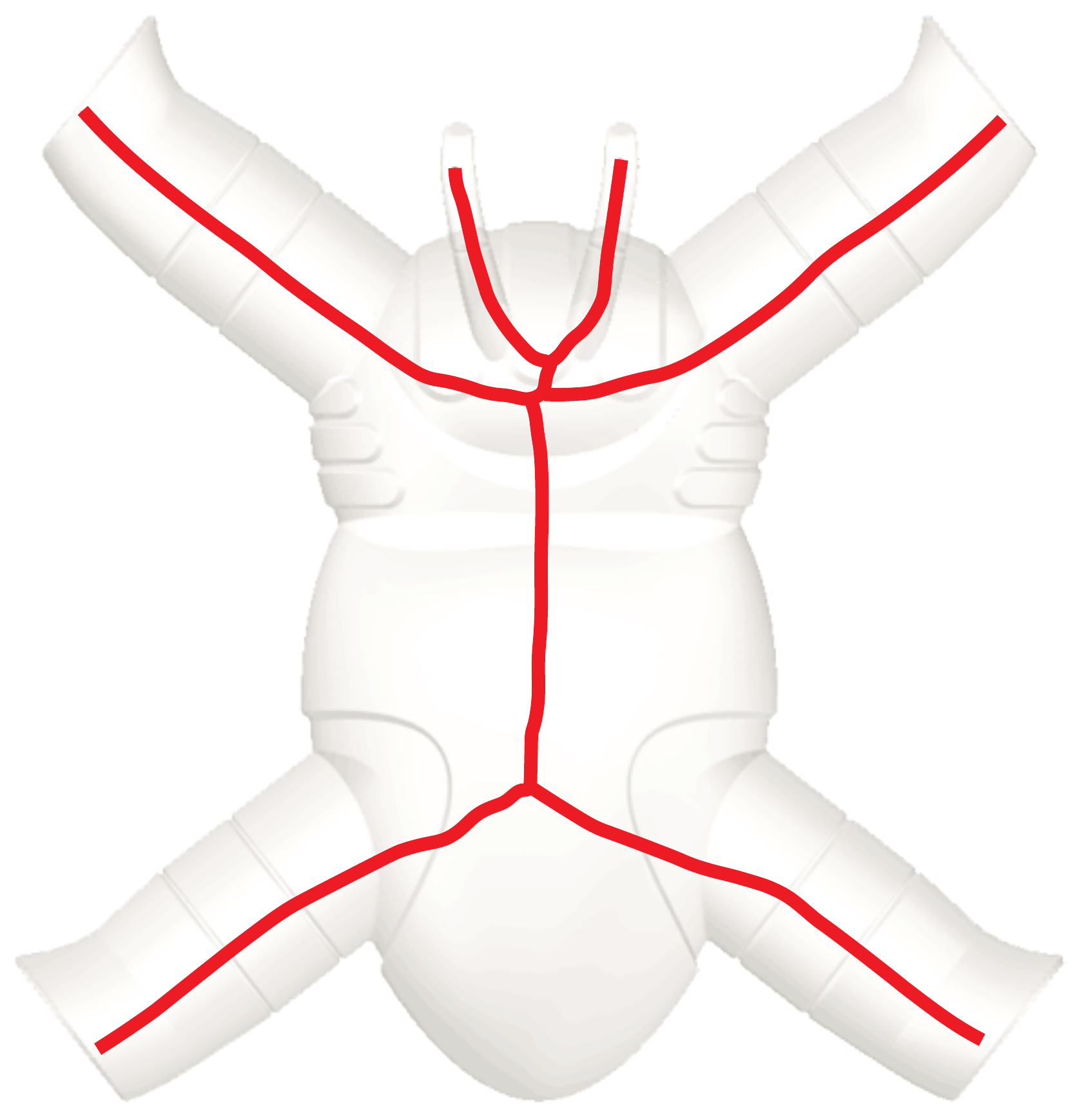}
\caption{The initial skeletal line of the beetle-like robot.}
\label{fig:beetle_skeleton}
\vspace{-10 pt}
\end{wrapfigure}

\begin{figure}[t]
  \centering
  \includegraphics[width=\linewidth]{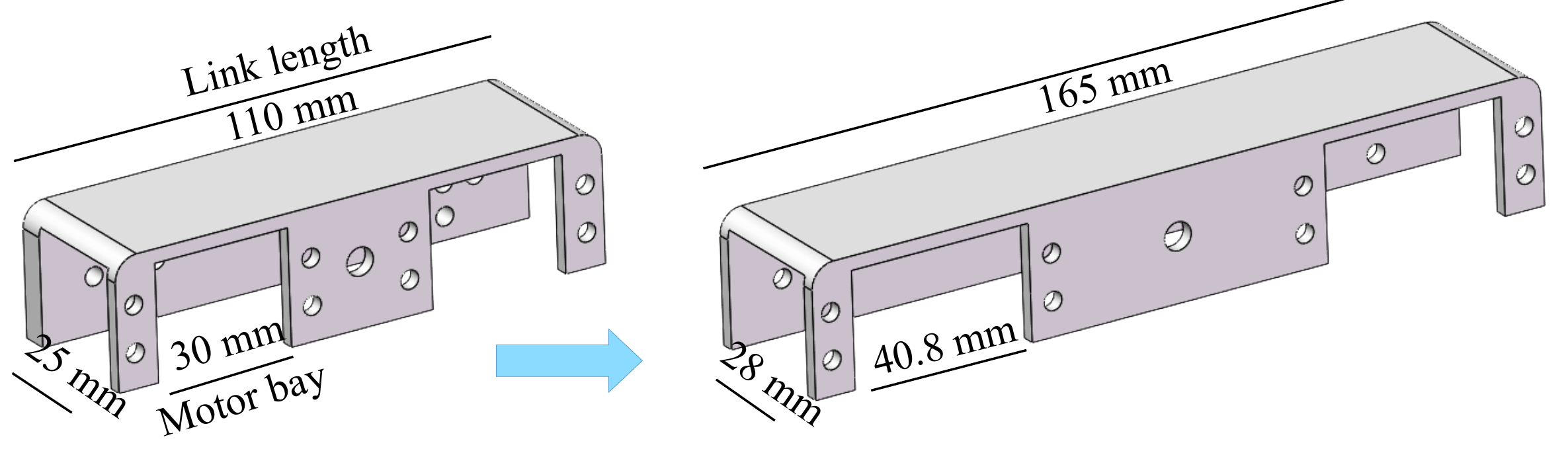}
  \caption{With the assistance of the developed \texttt{Solidwords} scripts, the user only needs to tweak semantic parameters like the link length, motor mount size, etc. to obtain a customized link. The size of all the pilot holes for screw installation remains unchanged under such edits.}
  \label{fig:link_editing}
\end{figure}

The design starts with a given 3D model that corresponds to the appearance of the robot. Our system extracts an initial skeletal line using the mesh contraction method~\cite{Au:2008:SEM} as shown in \figref{fig:beetle_skeleton}. This skeleton is actually an approximation of the medial axis of the model, and it is used as a general guide for the follow-up template embedding and editing. We employed the \emph{modular design} idea so that the user can edit the geometry of a template mechanical component to obtain a customized mechanical skeleton for quad-robots of various morphologies. To this end, several \texttt{SolidWorks} scripts are developed to assign semantic parameters
 of a link, such as the link length, motor mount size, etc., and the user only needs to tweak these intuitive parameters to obtain a personalized design without creating one from scratch. The size of pilot holes on the link for screw installation remains unchanged under such edits. An example is given in \figref{fig:link_editing}, where the lengths of the link and the motor bay are increased. Although a few iterations may be necessary during embedding, the developed \texttt{SolidWorks} scripts greatly accelerate the procedure.

\xwwedit{Typically, given a new surface model of a quad-robot, we embed limbs first and then adjust the geometry of the torso to make sure it fits the exterior skin. Specifically, a global scale of the mechanical structure template and local rotations of the links are first performed so that the template can be inside the input surface mesh. Then, the user can select the start and end points of a link on the extracted skeletal line and trigger the designed script to edit the link geometry to match the specified length and adjust the width of the link. Finally, the bracing unit of the torso is generated in a similar way as the skin creation(the details follows shortly), and we dig out holes to reduce its weight, for instance, the bracing unit for the torso of the beetle-like robot shown in the first picture in the second row of \figref{fig:teaser}.}

\begin{figure}[t!]
	\centering
	\includegraphics[width=\linewidth]{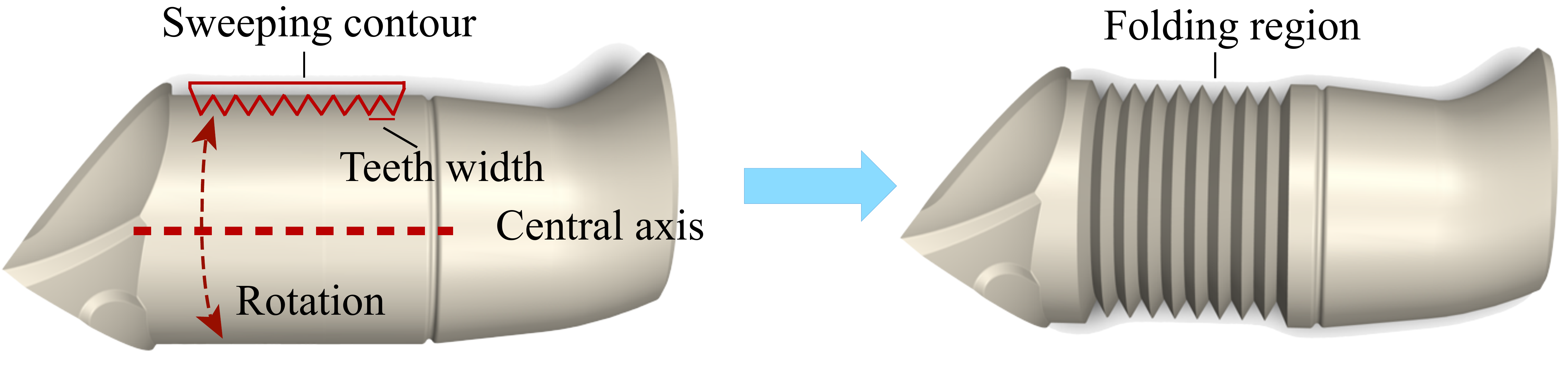}
	\caption{We add folding regions to facilitate the stretching deformation of the skin. The template folding region is similar to gear teeth, and it is formed by a sweeping cut operation, that is, the CSG difference between the volume surrounded by the original skin surface and the volume formed by rotating the sweeping contour along a central axis.}
	\label{fig:sweeping}
\end{figure}

\xwwedit{\noindent\textbf{Skin and folding regions creation}: The exterior skin of the robot is designed to be $8$ mm thick at the foot and $4$ mm thick at other parts by default, and it is created by the mesh hollow operation in \texttt{Materialize Magics}. This operation treats the space surrounded by the input 3D surface model as a solid and hollows the interior space of the solid to match the specified thickness parameters to create the skin that is amenable to fabrication.  When it is being bent, the skin can yield rather large resisting forces under stretching deformations. Regular commercial motors may not possess sufficient power to overwhelm the internal stretching. To resolve this practical issue, we add a few \emph{folding regions} on the original skin mesh, as shown in Fig.~\ref{fig:sweeping}. The folding region is created by applying a sweeping cut operation in \texttt{Solidworks} over the original skin surface where the motor is installed. This small treatment increases the skin area where substantial bending occurs and effectively reduces the resulting stretching force. Note that the hollow operation is performed after the creation of the folding regions. Fortunately, the two software are compatible in mesh file format.}

\subsection{Fabrication}
\label{sec:robot_fabrication}

\setlength{\columnsep}{5 pt}%
\begin{wrapfigure}{r}{0.35\linewidth}
	\vspace{-5 pt}
	%\begin{center}
	\includegraphics[width =\linewidth]{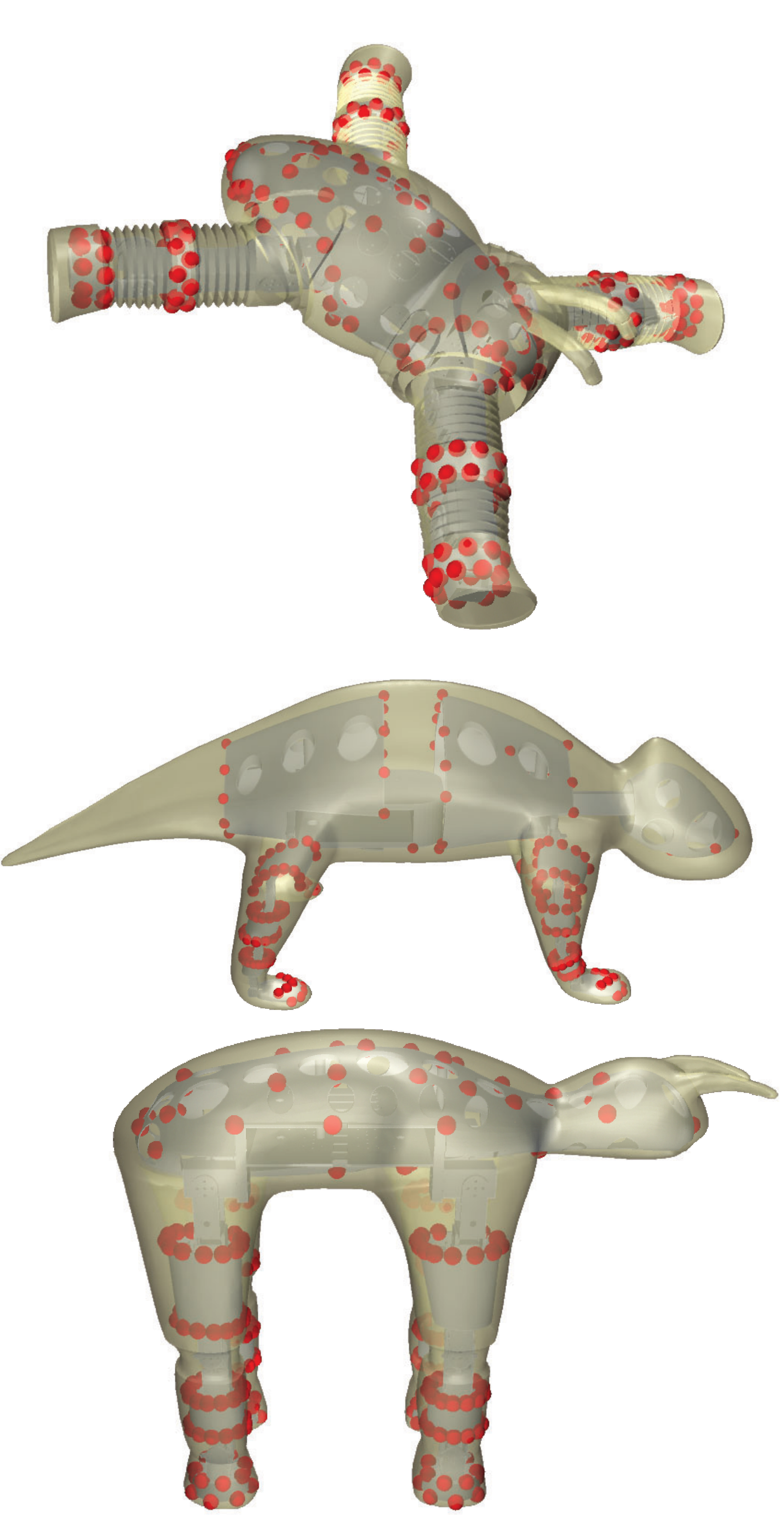}
	\caption{Glue vertices.}
	\label{fig:glue}
	\vspace{-10 pt}
\end{wrapfigure}

The mechanical structure of the robot is 3D printed with polypropylene-like stereolithography (SLA) resin, which is a widely used material for fabricating joints and low-friction moving parts. The exterior skin of the robot is made of a layer of soft rubbery material and fabricated via injection molding. To reduce the effort and cost of creating the skin molds, we fabricate the skin on a piece-by-piece basis: one limb has one skin piece, and the torso has two pieces as shown in \figref{fig:teaser} (Skin pieces). The skin-skeleton attachment is physically realized by another 3D printed bracing unit between the skin and the skeleton. This bracing unit is attached to each link on the skeleton and serves as a supporting structure between the rubbery skin and the mechanical skeleton (see \figref{fig:teaser} \& \figref{fig:glue}). \xwwedit{The purpose of this design is to expect that the friction between the skin and the printed parts can disable the relative motion between skin and skeleton at these parts, which is verified in the physical validation. We thus select the glue vertices in \figref{fig:glue} according to the position of skin-skeleton attachment parts so that the coupling constraints can reflect this physical setting.} The mass matrix and the inertia tensor of this bracing unit are integrated in our multibody subsystem dynamics. Finally, skin pieces are glued together after all skin pieces are installed using nonreactive PVA adhesive.

\vspace{5 pt}
\noindent\textbf{Motor specification}\hspace{5 pt}
We use the \texttt{MG995R} servo motor to drive the motion of the skinned robot. The motor's size is $40.8\times20\times38$ mm with the maximum torque of $20~$ kg$\cdot$cm under $6.4$ V (i.e. $U_m=1.96$ N$\cdot$m in Eq.~\eqref{eq:original_optimization}). In total, 12 motors are installed in the beetle-like robot. All the motors are controlled with an \texttt{Arduino} board, which supports up to 32 motors.

\section{Experimental Results}
\begin{figure*}[t!]
  \centering
  \includegraphics[width=\textwidth]{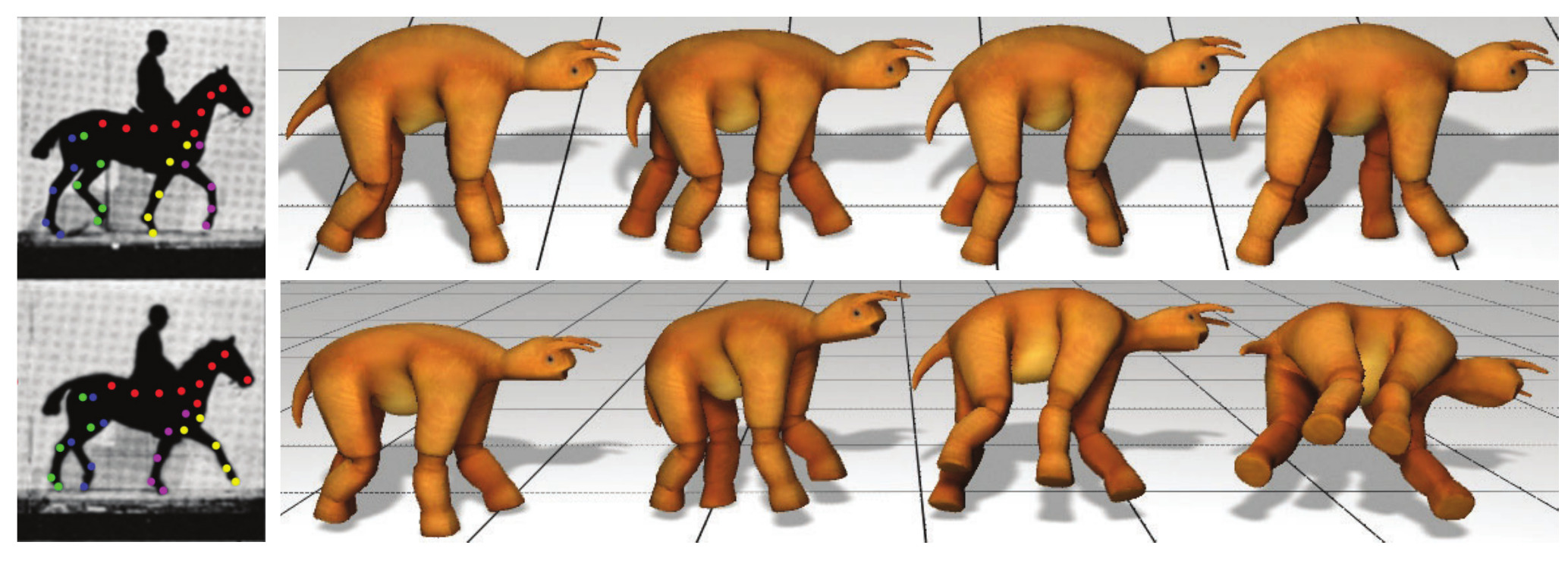}
  \caption{Trotting motion for the monster-like robot. (a) Marked joint positions. The color of a dot indicates to which part of the horse skeleton it belongs, and the positions are mapped to our monster skeleton joint angles using the space-time optimization with only kinematic constraints.  (b) The designed trotting motion with our alternating algorithm. (c) The unstable motion simulated by the frame-by-frame optimization when the skin deformation is not considered in the space-time optimization. }
  \label{fig:monster_fastmotion}
\end{figure*}

\begin{figure*}[t!]
  \centering
  \includegraphics[width=\textwidth]{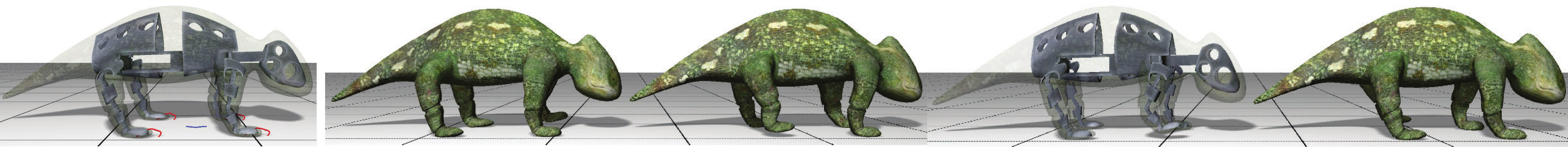}
  \caption{The motion of a lizard-like robot. We edit the second template in \figref{fig:templates} with \texttt{SolidWorks} scripts to create the design of its mechanical skeleton. With user provided inputs, our system generates plausible motions of this robot.}
  \label{fig:lizard}
\end{figure*}

In this section, we first report the torque limits and folding region experiments in the motion design of the beetle-like robot to validate its physical feasibility. The mechanical skeleton of this robot is designed based on the first template in \figref{fig:templates} and fabricated using 3D printing. A comparison to kinematic optimization only is also provided for this robot. Second, we report the motion design results for two additional quad-robots: a monster-like robot and a lizard-like one. The performance of our optimization algorithm depends on the number of vertices on the skin mesh, the number of glue vertices, and the number of joints of the mechanical skeleton. The frame interval $\Delta t$ is $0.005$ second, and we employ the discrete collision detection algorithm to handle self-collisions and foot-ground collisions. Normally a motion cycle has around $500$ frames. Our optimization algorithm was implemented on a desktop PC with an \texttt{intel} \texttt{i7-7700} CPU and 16 GB memory. The soft skin is made of isotropic rubber material whose Young's modulus is $0.09~GPa$, and Poisson's ratio is $0.46$. \tabref{tab:timing_statistics} reports some essential physical and simulation statistics of these three examples. The space-time optimization step with approximated skin deformation for a skeleton is around 40 seconds. For the slow-walking motions as shown in \figref{fig:teaser} and \figref{fig:lizard}, we only need  one iteration to converge, since the frame-by-frame optimization can reproduce the motion from space-time optimization step well with physical constraints. For the relatively fast trotting motion in \figref{fig:monster_fastmotion}, the algorithm converges after two iterations.
\begin{table}[t!]
\begin{center}
\begin{tabular}{|c|c|c|c|c|c|}
\hline
Robot & Joints & Skin &Glue & Weight & Opt. \\
\hline
Beetle-like & $12$ & $14,152$ & $140$ & $3.9kg$ & $\sim3.2~s$ \\
\hline
Monster-like & $16$ & $16,624$ & $301$ & $11.87kg$ & $\sim8.98~s$\\
\hline
Lizard-like  & $17$ & $18,261$  & $176$ & $10.8kg$ & $\sim8.41~s$ \\
\hline
\end{tabular}
\end{center}
\caption{Physical and simulation statistics of three tested robots. Joints: the number of joints on the skeleton. Skin: the number of vertices on the skin mesh. Glue: the number of glue vertices. Weight: the physical weight of the robot. Opt.: the average time used for optimizing Eq.~\eqref{eq:original_optimization} of one motion frame.}
\label{tab:timing_statistics}
\vspace{-15pt}
\end{table}

\begin{figure}[t!]
  \centering
  \includegraphics[width=\columnwidth]{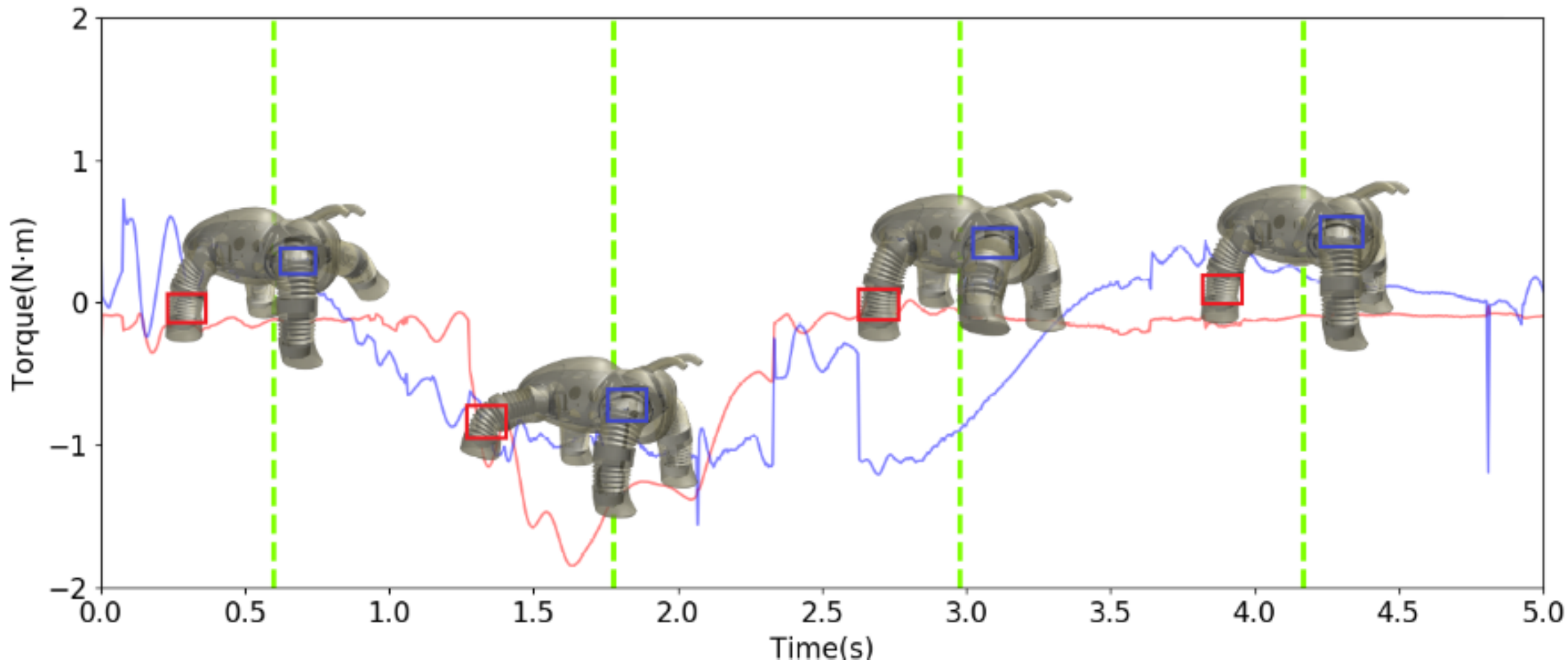}
  \caption{We plot the torque values of the optimized motion planning of the beetle-like robot. Orange curve: the torque curve at an ankle joint (at the orange box). Blue curve: the torque curve at a shoulder joint (at the orange box).}
  \label{fig:torque_curve}
\end{figure}

%The mechanical skeleton of beetle-like robot shown in \figref{fig:system_flowchart} is designed according the first template in \figref{fig:templates}, and the robot is fabricated out to verify its physical feasibility. In this section, we report experimental results on motion plan generation and the physical verification on the beetle-like robot. Timing and statistics are reported in \tabref{?}.

\noindent\textbf{Torque limit}:\hspace{5 pt}
The torque limit in the motion optimization (i.e. Eq.~\eqref{eq:original_optimization}) of the beetle-like robot is set as $1.96$ N$\cdot$m to match the physical torque limit of the used \texttt{MG995R} motor. To verify if this hard constraint is faithfully enforced during the optimization, we examine torque values that are calculated by our motion design system after per-frame optimization. The result is reported in \figref{fig:torque_curve}, where torque curves at an ankle joint and a shoulder joint are plotted. It can be seen that the imposed motor constraint successfully bounds the torque magnitude to be within the limit to ensure that the designed motion is physically possible.

\begin{figure}[t!]
  \centering
  \includegraphics[width=\columnwidth]{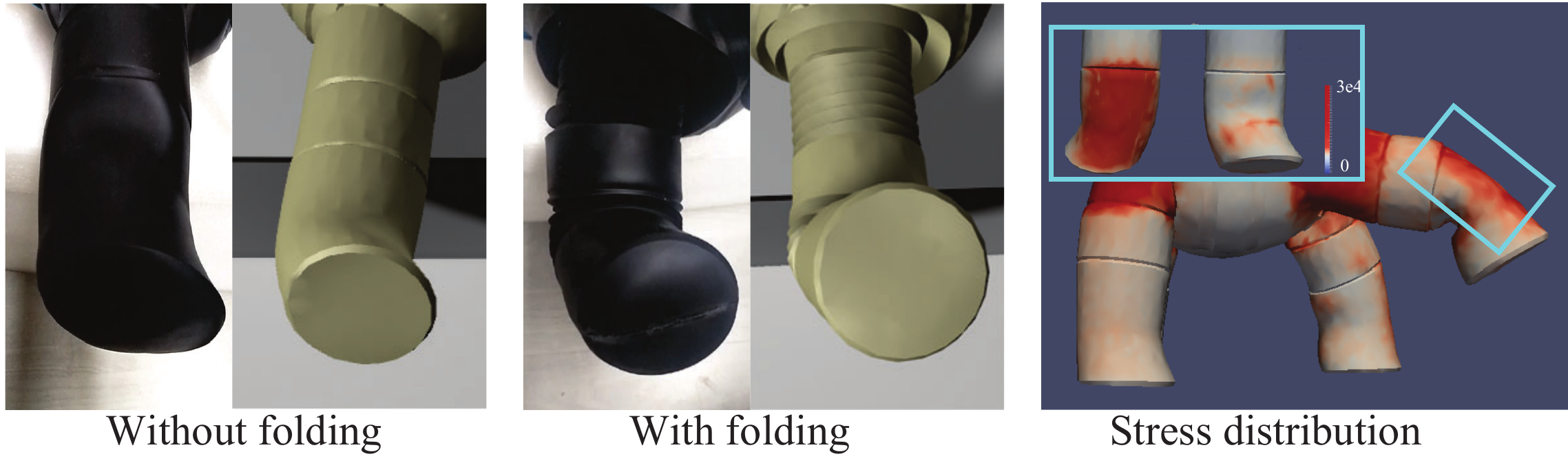}
  \caption{Adding folding regions significantly relieves the stretching stress over the skin. We program the motor at an ankle joint of the beetle-like robot to rotate $\pm 70^\circ$ within 2 seconds in this experiment. A smooth skin can only be bent around $\pm 20^\circ$, while the folded skin is able to reach the desired deformation. The physical experiment (black skins) results are consistent with simulation results (yellow skins).}
  \label{fig:fold_region}
\end{figure}
\vspace{5 pt}
\noindent\textbf{Folding regions}\hspace{5 pt}
Adding folding regions to the robot's skin is an effective fabrication artisanry to enable the robot assembly using off-the-shelf servo motors and lower the fabrication cost. To demonstrate its advantage, we compare the skin deformation under the joint rotation when the robot is attached to a regular soft skin and a folded skin. Both skins are fabricated using the same materials. It can be seen from \figref{fig:fold_region} that rotating joints yield large stretching stress over the skin, which could easily go beyond the physical capacity of many commercial servo motors. In this test, we follow the aforementioned motor specification by setting the maximal torque to $1.96$ N$\cdot$m and test if this power is sufficient to generate the necessary skin deformation. We set our target bending angle to $\pm70^\circ$, which is a common value in many walking gaits. The motor is programmed to reach this target in $2$ seconds. Our simulation shows that the smooth robot skin without folding region prevents the motor from producing sufficient joint rotation and the maximum angle that can be reached is only about $\pm20^\circ$. With the folding region, swept by an $8$ mm-depth tooth over the smooth skin, our simulation predicts that the motor is able to generate the desired rotation. The physical experiment results are quite consistent with our simulation prediction as reported side by side in \figref{fig:fold_region}.

\begin{figure}[t]
\vspace{-5 pt}
\includegraphics[width =\linewidth]{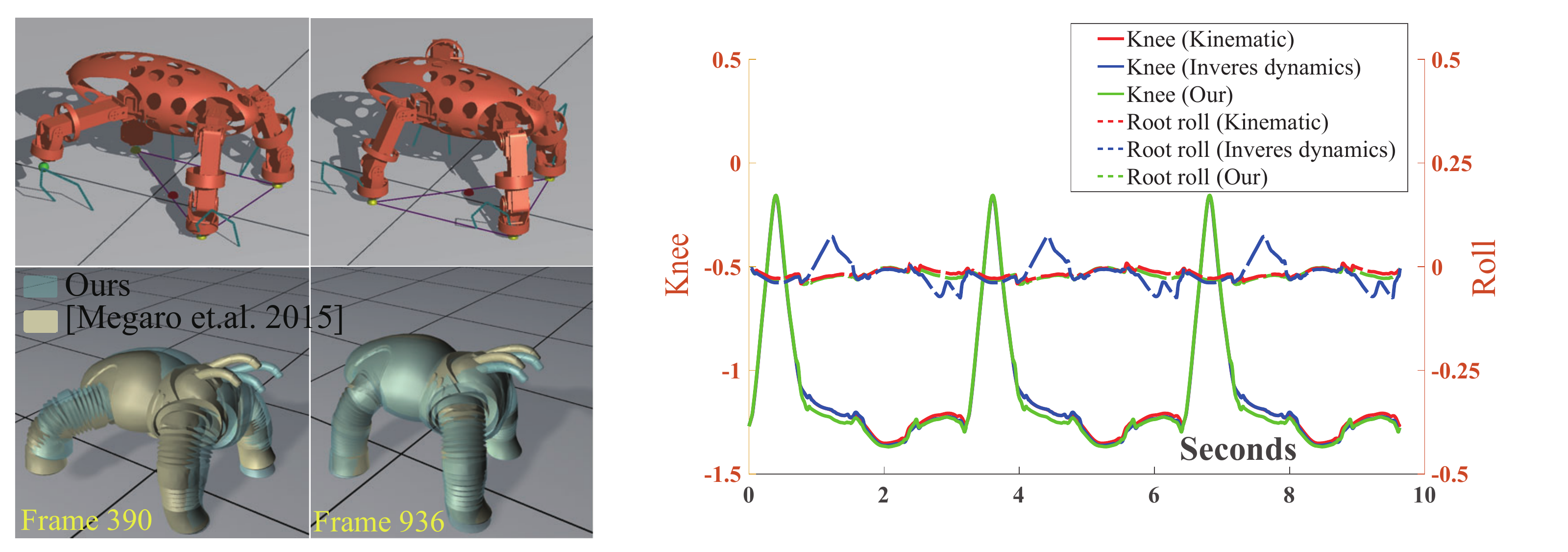}
\caption{The motion plan generated by the kinematic-only optimization~\protect\cite{Megaro:2015:IDR} leads to unstable walking sequences. Left: Selected frames. Right: The joint angle curves. Kinematic: the kinematic optimization result. Inverse dynamics: the simulation result by following the kinematic optimization result. Our: our optimization result. The large roll angles of the root link is due to the unstable pose using  kinematic-only optimization. Please see the accompanying video for the full comparison.}
\label{fig:comparison}
\vspace{-5 pt}
\end{figure}
%\vspace{5 pt}

\begin{figure}[t!]
  \centering
  \includegraphics[width=\columnwidth]{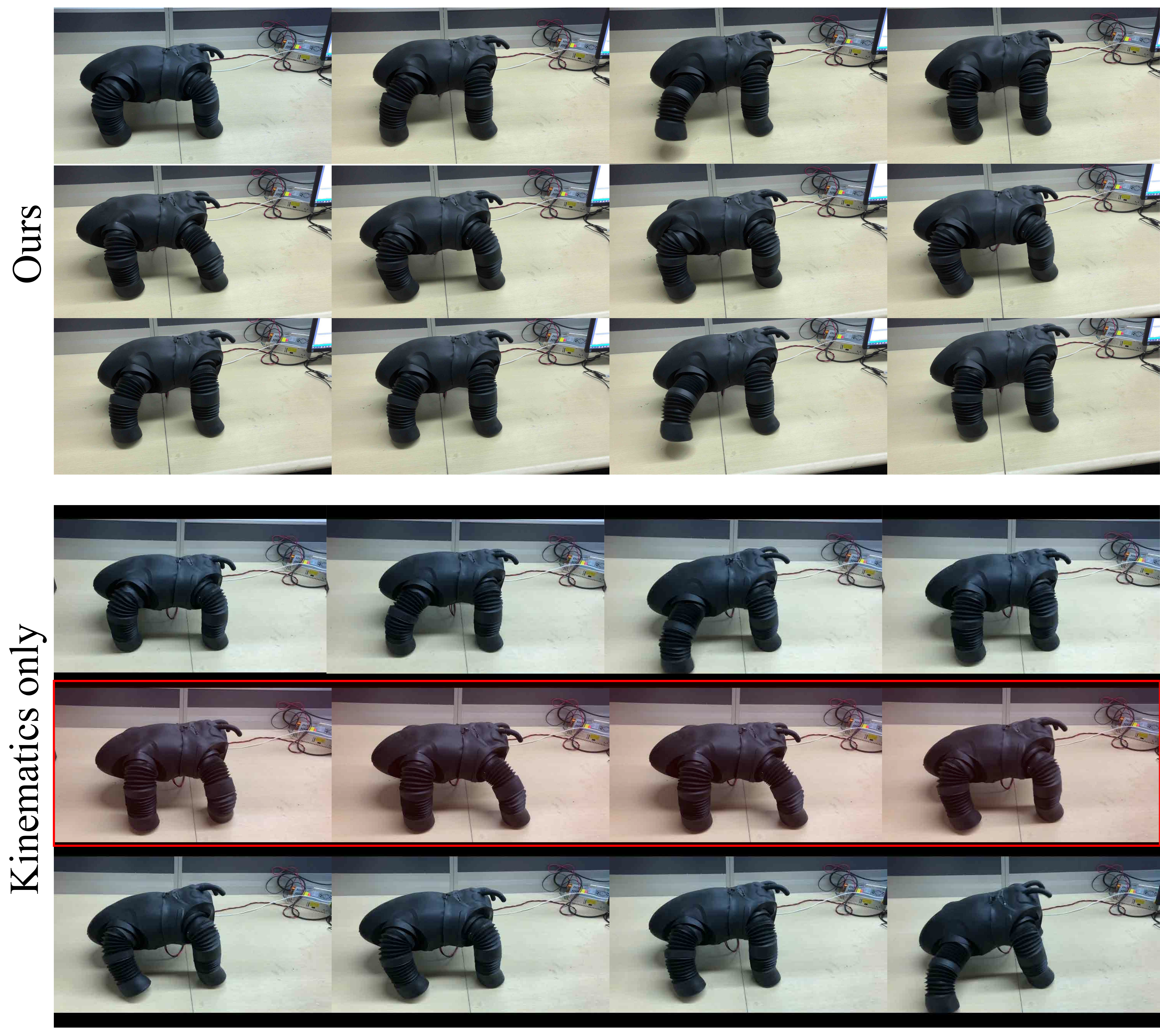}
  \caption{Physical experiments show that kinematic-only optimization is not a feasible solution for skinned robots -- there are noticeable back steps (highlighted with a red box) in a motion cycle as the driving torques, after damped by the skin deformation, are not strong enough to produce necessary normal contact forces. Please refer to the accompanied video for a clearer comparison.}
  \label{fig:physical_comparison}
\end{figure}

\noindent\textbf{VS. kinematic-only optimization}:\hspace{5 pt}
In contrast to robots with only rigid mechanical skeletons, skinned robots exhibit a much more complicated dynamic behavior, which should be fully incorporated during the motion design. To illustrate the necessity of incorporating influences of the soft skin, we compare the motion plans generated using our method and the one by Megaro et al.~\cite{Megaro:2015:IDR}. Because the primary focus in~\cite{Megaro:2015:IDR} is to design robot creatures with only rigid links, Megaro and colleagues used a kinematic-based optimization strategy, which includes the trajectories of COM, COP, end effectors, and the footfall pattern. Based on the resulting motion plan, we compute the corresponding driving torques at joints using inverse dynamics. Specifically, the driving torques are computed by imposing another set of rotation constraints over the skeleton in Eq.~\eqref{eq:system_equation} using the Lagrange multipliers method (with necessary complimentary constraints and inequality constraints to handle the ground contact and motor torque limit). The constrained joint rotation corresponds to the one obtained from the kinematic-based motion plan, and the multipliers represent required generalized constraint force, which are converted to joint torque via $\mathbf{J}_{\omega k}$ to achieve the target joint rotation. As shown in \figref{fig:comparison}, the physical simulation results suggest that a kinematically valid joint trajectory does not guarantee a smooth walking cycle of the skinned robot, even though the constraints of COM/COP are also specified in the kinematic optimization without skin information. The coincidence, for the knee joint, of the joint angle curves of kinematic and inverse dynamics shows that our inverse dynamics computation can track the kinematic motion plan well, and the roll angle of the root link experiences a larger variation in the inverse dynamics simulation. This is also verified in the physical experiment as shown in \figref{fig:physical_comparison}. The motion plan obtained using only kinematic optimization leads to a shaky motion. We also observe backward motions as highlighted in the figure. Our method, because it fully considers various physics conditions and constraints, yields a much better result.

\begin{figure}[t]
  \centering
  \includegraphics[width=0.95\columnwidth]{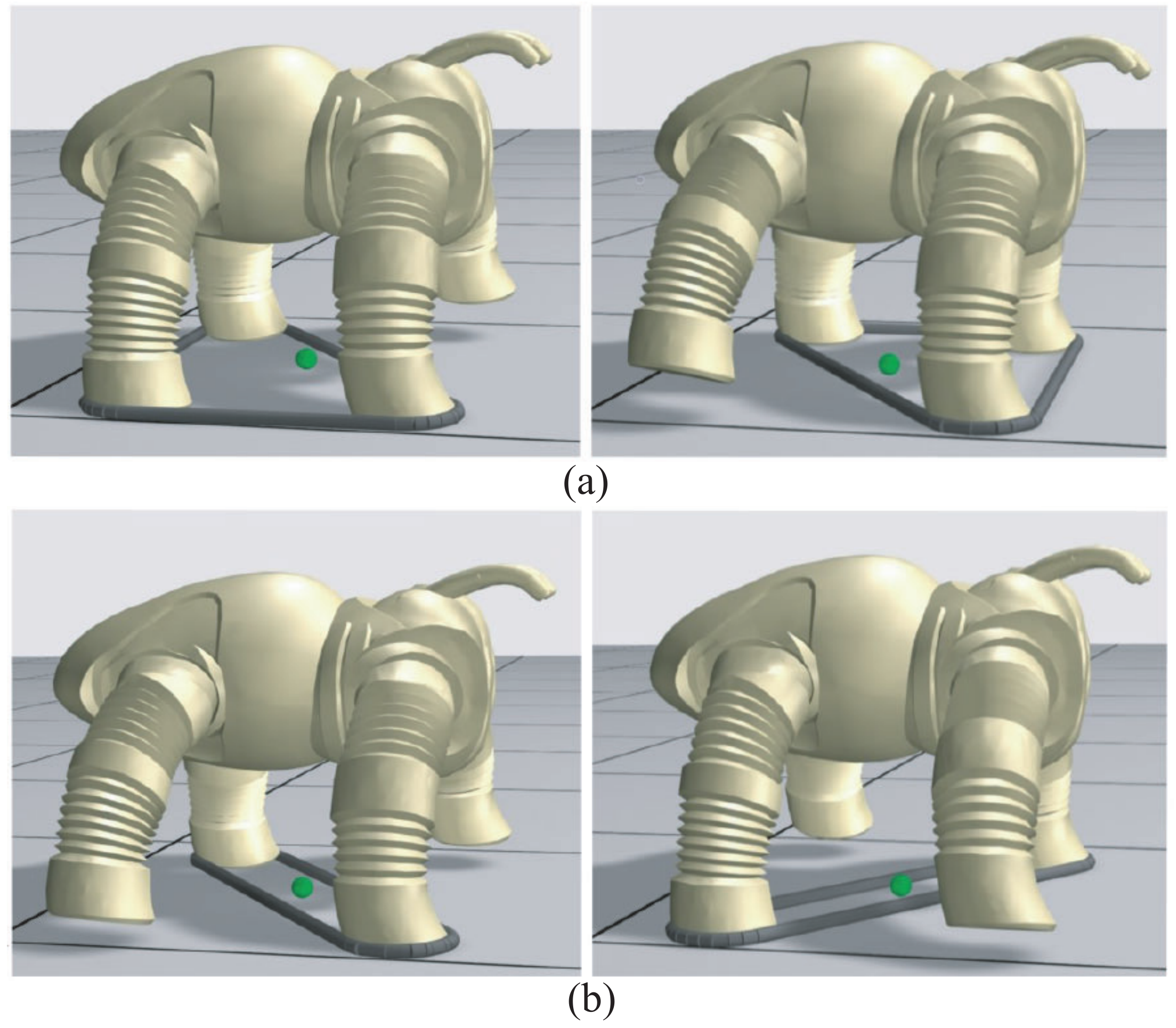}
  \caption{\xwwedit{The foot lifting motion for the Beetle-like robot. (a) Single-foot lifting. (b) double-foot lifting. The green balls indicate the COP positions and gray lines the support polygons. Our optimization algorithm can constrain the COP to be inside the support polygon.}}
  \label{fig:COP_TEST}
\end{figure}
%\vspace{5 pt}

\begin{figure}[t]
  \centering
  \includegraphics[width=0.95\columnwidth]{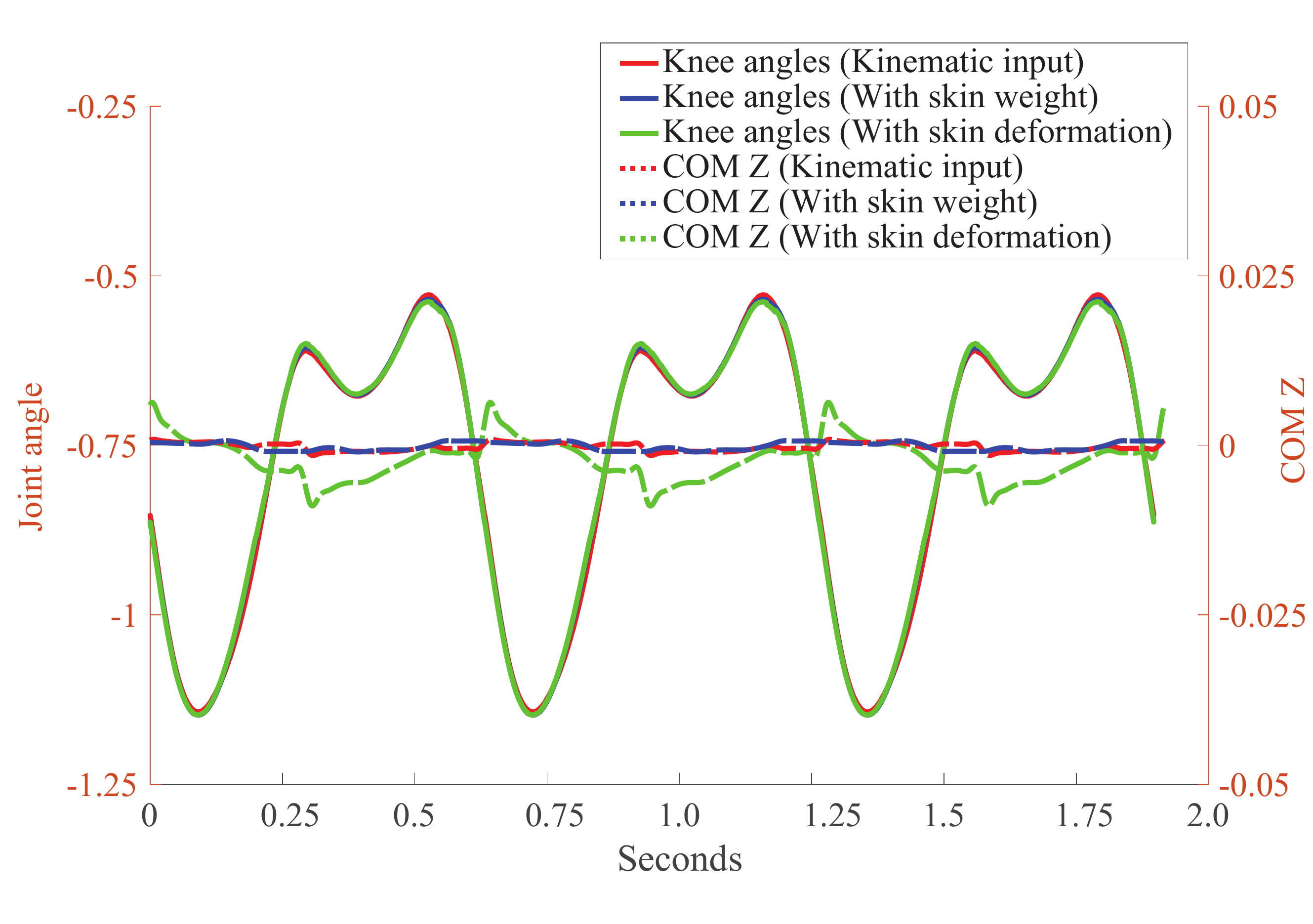}
  \caption{The comparison of joint angle curves and COM positions. With skin weight: the curves are from the initial space-time optimization result with only skin weight.  With skin deformation: curves from the space-time optimization result with skin deformation simulated by frame-by-frame optimization. COM Z: the $z$ component of the COM, representing the COM movement from left to right during the motion.}
  \label{fig:space_time_opt_result}
\end{figure}

\begin{figure}[t]
  \centering
  \includegraphics[width=0.95\columnwidth]{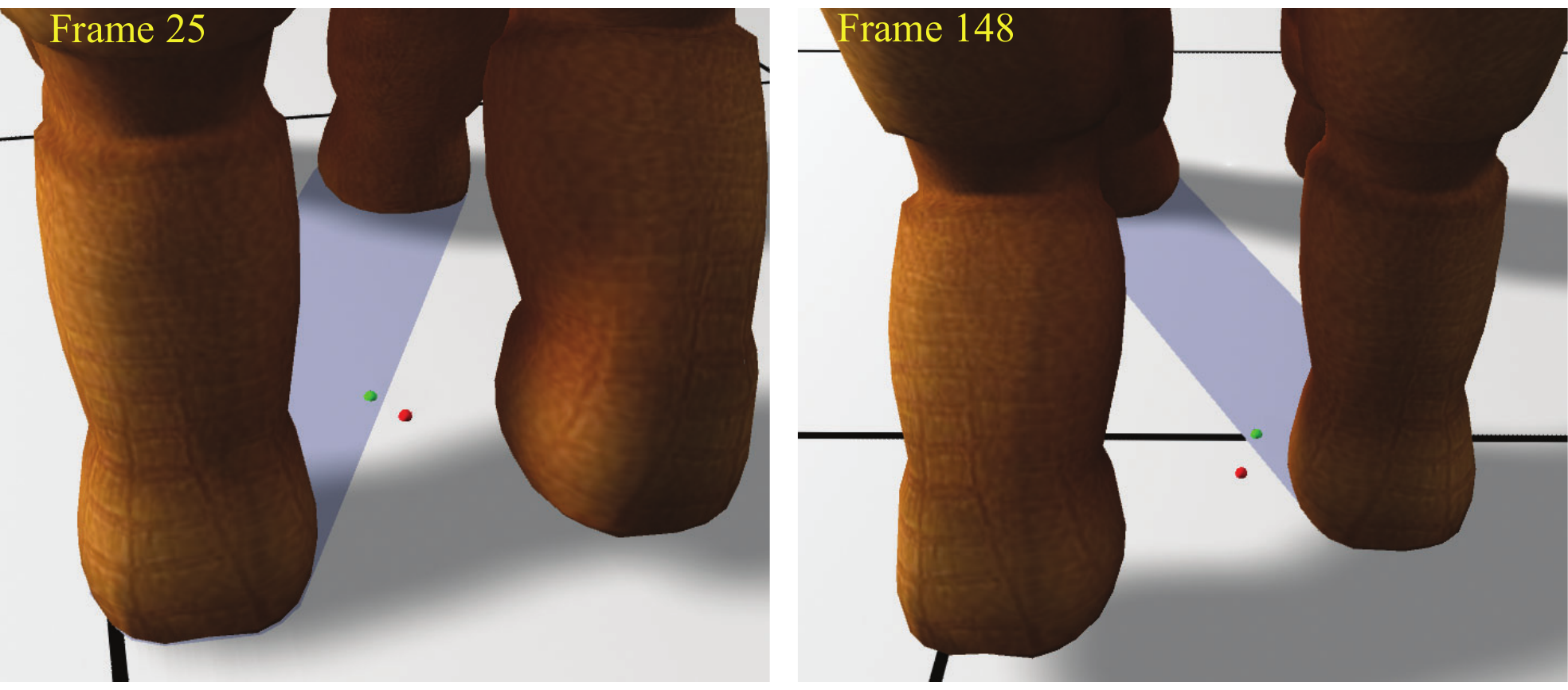}
  \caption{The effect of the COP constraints in the trotting motion plan. Red balls: The COP positions computed using the motion plan after initialization. Green balls: The COP positions optimized with the skin mesh deformation. The supporting polygons are in cyran.}
  \label{fig:COP}
\end{figure}
\vspace{5 pt}

\begin{figure*}[t!]
  \centering
  \includegraphics[width=\textwidth]{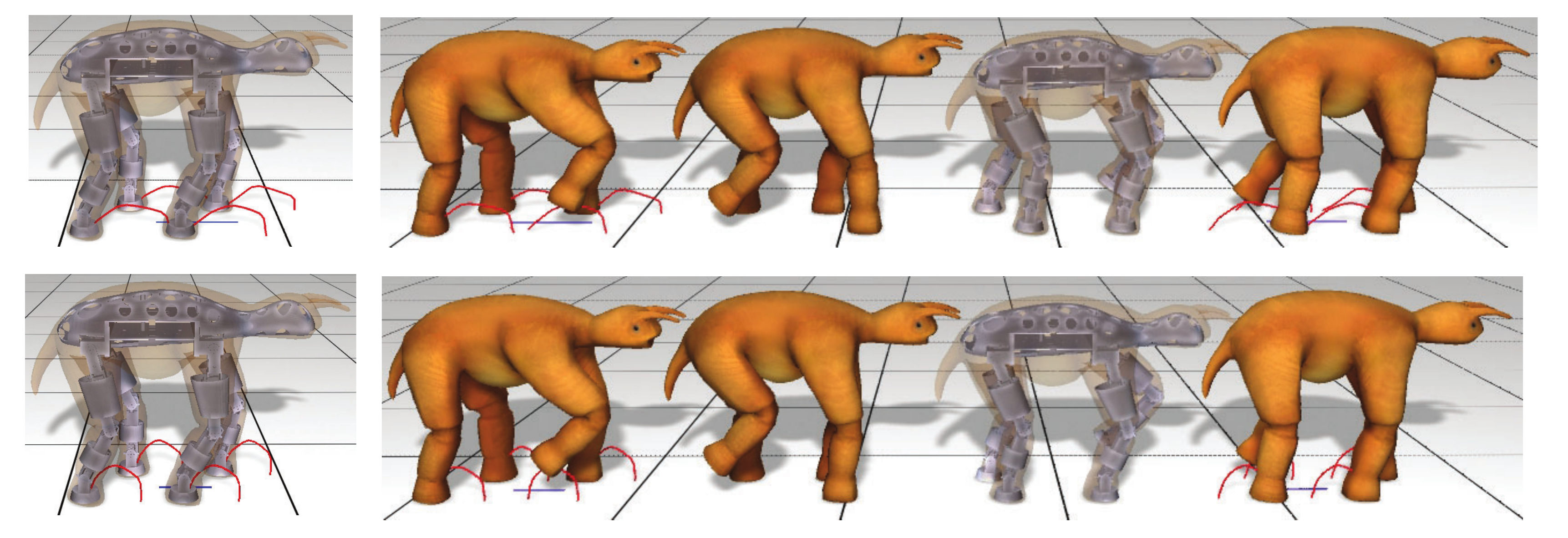}
  \caption{The monster-like robot takes two different input foot trajectories, and our system computes natural and physically correct motions for both inputs.}
  \label{fig:monster_result}
\end{figure*}

\vspace{5 pt}

\noindent\textbf{Motion design results}:\hspace{5pt}
\xwwedit{\figref{fig:COP_TEST} illustrates the simple foot lifting motions designed by our system for the beetle-like robot. These two motions, i.e., single-foot lifting and double-foot lifting, are also used to show the COP is constrained to be inside the support polygon with our optimization algorithm. Please see the accompanying video for the animation.}

The embedded skeleton of the monster-like robot shown in \figref{fig:templates} is designed using the third template. The weight of this robot is $11.87~kg$, and its size is $48.5$ cm $\times$ 64.6 cm $\times$ 27 cm. The young modulus at the tail and belly of the monster is reduced by 85\% to demonstrate the dynamics of the skin.  Two different input foot trajectories are used to generate the walking motions for the robot. As shown in the leftmost column in \figref{fig:monster_result}, the first trajectory has a longer stride length but lower step, while the second one has shorter stride and higher step. Our system is able to accommodate such variations and produces a smooth and physically correct motion plan. The walking speed for these two walking motions are $0.11$ meter/second and $0.06$ meter/second.

We generate a trotting motion of $0.45$ meter/second speed for the monster-like robot to demonstrate that our system can support fast motion. In this example, the trajectories of joint positions are labelled using the horse motion pictures photographed by Eadweard Muybridge, a famous photographer for his work on motions. The joint positions are mapped to a horse motion with the specified speed using the method in~\cite{HuangCASA:RHGS} and re-targeted to the skeleton of our monster. This initial kinematic motion (please see the accompanying video for the motion) is then optimized using our alternating motion optimization algorithm to turn it into a physically feasible motion.

We notice that the flighting phase of the initial kinematic motion is not consistent with the foot contact plan. To be more specific, the time of the flighting phase is not enough for the monster to return back to the ground. Thus, the space-time optimization with physical constraints, especially the momentum constraint, is necessary to eliminate such inconsistency, which is critical to the success of the alternating optimization. \figref{fig:COP} illustrates the effect of the COP optimization. The red balls indicate that the COP positions at two frames in the initialized trotting motion plan are outside of supporting polygon after the first space-time optimization that does not account for the deformation of the skin mesh. Thus, the frame-by-frame simulation fails to produce a stable trotting motion with its initially optimized motion plan, as shown in the second row of \figref{fig:monster_fastmotion}. The COP constraint is turned off to produce this failed example once this constraint can not be satisfied by the solver. After the second iteration, the COPs are moved into supporting polygons, as indicated by the green balls. A smooth trotting motion can then be generated as shown in the first row on the right of \figref{fig:monster_fastmotion}. The comparisons of the optimized joint angles and the $z$ components of COP are illustrated in \figref{fig:space_time_opt_result}. The variation of $z$ components indicates the COP moves from left to right so that it is inside the supporting region.

%Similarly, if the fast vibration of the monster belly is simplified to be static in the space-time optimization, the frame-by-frame optimization will also fail to produce a feasible motion. As a result, the accumulation of the joint angle tracking errors will interfere with the footfall pattern, as shown in the second row of the \figref{fig:monster_fastmotion}, and the frame-by-frame simulation fails to produce a smooth trotting motion.

%To produce a fall-down simulation result in this case, the COP constraint is disabled if we detect that the per-frame optimization fails. Please see the accompanied video for the animation.

Another example is reported in \figref{fig:lizard}. The mechanical skeleton of this robot is further edited based on the second template of  \figref{fig:templates}. We lengthened the torso and shortened the limbs to fit this template into the input model ($125$ cm $\times$ 47 cm $\times$ 46 cm). Our system also produces plausible motion plans for this quad-robot.

%The torque limit in the motion optimization (i.e. Eq.~\eqref{eq:original_optimization}) of the beetle-like robot is set as $1.96~N\cdot m$ to match the physical torque limit of the used \texttt{MG995R} motor. To verify this hard constraint is faithfully enforced during the optimization, we examine toque variations at all the joints that are calculated by our system during the locomotion. The result is reported in \figref{fig:torque_curve}, where torque curves at an ankle joint and a shoulder joint are plotted. It can be seen that even though the calculated torques frequently hit the cap, the imposed motor constraint successfully bounds the torque magnitude to be within the limit.

\noindent\textbf{Failure case}:\hspace{5pt} While our system is stable in the generation of slow walking motion, we find the generation of fast trot motion is sensitive to the physical parameters. When the mass of the monster is increased to two times, its influence on the mass center cannot be balanced in the optimization and the COP constraint is violated in the space-time optimization result, possibly due to its conflict between the foot contact constraint. Hence, the frame-by-frame optimization will fail to produce a stable motion. Such situation might be handled through the integration of foot plan sampling step in~\cite{Wampler:2009:OGF:1531326.1531366}.

\section{Conclusion and Future Work}
In this paper, we have presented a fabrication-oriented motion planing algorithm and detailed design/fabrication procedures for personalized skinned quad-robots. The physical constraints, such as the equations of motion of the skinned robot and the motor constraints, are integrated into the motion planning such that the resulting motion plan is physically and dynamically feasible. The condensation formulation allows us to conveniently establish the nonlinear relationship between external forces and the target kinematic parameters of the locomotion and to reach a QPCC formulation for the motion design. Our experiments show that the system is able to assist regular users to obtain natural and smooth motions designed for skinned quad-robots.
%The beetle-like robot also is physically fabricated and tested.

%We have tested our system using both simulated robots and fabricated ones

In the future, we want to explore a gait synthesis algorithm to generate the motion plan from high-level parameters, such as velocity and turning angles. Combining captured gait data and optimization with dynamic constraints has the potential to significantly reduce users' labor efforts of creating such motion planning. Currently, the coupling between the FEM simulation of the soft skin and the rigid body dynamics of the mechanical structure is not fast enough for a closed-loop control of skinned quad-robots. We want to explore model reduction or homogenization techniques to reduce the computational cost of FEM simulation and produce interactive feedback to control skinned robots online.

%many existing contributions simplify the problem using one-way rigid-to-soft coupling -- the articulated rigid skeleton triggers secondary %inertial dynamics of the skin, yet the influence of deformable skin dynamics is ignored~\cite{capell2002interactive,lee2009comprehensive}. %Liu and colleagues~\cite{liu2013simulation} handled the influence from soft skin to the rigid skeleton. However because they used ODE %as a black box to simulate the skeleton, the dynamics of both subsystems are decoupled at different time steps, and each subsystem is %updated \emph{separately}. That is the configuration of the rigid skeleton is computed based on the information of the soft skin at the %previous time step. Departing from existing methods, we aim to provide an \emph{accurate} predication of the skinned robot to the user. %Therefore, we integrate both multibody system and deformable simulation in a fully two-way coupled manner and we solve all the unknown DOFs %from both subsystems simultaneously. As elaborated later in this section, we use Lagrangian mechanics for both rigid skeleton and soft skin %and obtain a symmetric formulation for these two subsystems. We use the Euler angle to parameterize the local joint rotation to avoid %excessive external hard constraints imposed. Rather than widely-adopted co-rotational elasticity, the soft skin is modeled as a Neo-Hookean %solid~\cite{ogden1972large} in order to better incorporate large local deformation induced by joint rotation. Self-collision is also fully %accommodated in our system.

\appendices
\section{The Equations of Motion}
\textbf{Lagrangian multibody dynamics}: The multibody rigid skeleton of the robot is a kinematic tree of links connected by joints. Its classic Lagrangian mechanics formulation can be found in ~\cite{shabana2013dynamics}, where the DOFs of the skeleton are specified as the the generalized coordinate $\mathbf{q}$ of the joint angles and the root translation.
%\begin{equation}\label{eq:lagranian}
%\displaystyle\frac{\mathrm{d}}{\mathrm{d}t}\left(\frac{\partial \mathcal{L}}{\partial \dot{q}_i}\right)-\frac{\partial\mathcal{L}}{\partial %q_i}=0.
%\end{equation}
%$\mathcal{L}$ is the \emph{Lagrangian}, defined as $\mathcal{L}=T-V$ for the total kinetic energy $T$, and the system's potential energy %$V$. Expanding Eq.~\eqref{eq:lagranian}

The equation of motion for the articulated rigid skeleton can be written as
\begin{equation}\label{eq:eom_rigid}
\mathbf{M}_r(\mathbf{q})\ddot{\mathbf{q}}+\mathbf{D}_r\dot{\mathbf{q}}+\mathbf{f}_r(\mathbf{q},\dot{\mathbf{q}})=\mathbf{g}_r,
\end{equation}
where the subscript $[.]_r$ denotes variables for the skeleton rig. Detailed derivation of $\mathbf{M}_r$ and $\mathbf{f}_r$ in Eq.~\eqref{eq:eom_rigid} can be found in the excellent tutorial by Liu and Jain~\cite{liu2012quick}. $\mathbf{D}_r$ is the damping matrix. We refer to $\mathbf{M}_r$ as the \emph{rigid mass matrix} in order to differentiate it from the mass matrix of the soft skin. For a skeleton with $K$ links,
\begin{equation}\label{eq:rigid_mass}
\mathbf{M}_r=\sum_{k=1}^K\mathbf{J}^\top_k
\mathbf{M}_{ck}
\mathbf{J}_k,\quad\text{and}\quad \mathbf{M}_{ck}=
\left[
\begin{array}{cc}
m_k\cdot\mathbf{I} & \mathbf{0}\\
\mathbf{0} & \mathbf{I}_{ck}
\end{array}
\right],
\end{equation}
where $\mathbf{I}$ is the identity matrix, $\mathbf{I}_{ck}$ is the inertia tensor for the $k$\textsuperscript{th} rigid link, and $m_k$ is the mass of the link.
%\bernd{subscript k is missing in the formula above?!?}
The Jacobian matrix $\mathbf{J}_k = \left[\mathbf{J}_{vk}^\top, \mathbf{J}_{\omega k}^\top\right]^\top$, where $\mathbf{J}_{vk}=\partial \mathbf{x}_k/\partial \mathbf{q}$, relates the Cartesian coordinate $\mathbf{x}_k$ of the link's COM to the generalized coordinate $\mathbf{q}$. Similarly, $\mathbf{J}_{\omega k}$ relates the angular velocity $\pmb\omega_k$ to the generalized velocity $\dot{\mathbf{q}}$ such that $\pmb\omega_k=\mathbf{J}_{\omega k}\dot{\mathbf{q}}$. It is noteworthy that the rigid mass matrix is not constant because of the orientation-dependent inertia tensor and the Jacobian matrix.

Non-inertia forces like Coriolis and centrifugal forces that couple the generalized coordinate are included in $\mathbf{f}_r$.
%represented as $\mathbf{f}_r$, which can be written as:
%\begin{equation}\label{eq:rigid_internal}
%\mathbf{f}_r=\mathbf{C}(\mathbf{q},\dot{\mathbf{q}})\dot{\mathbf{q}}=\sum_{k=1}^K\left(
%\mathbf{J}^\top_k
%\mathbf{M}_{ck}
%\dot{\mathbf{J}}_k
%+
%\mathbf{J}^\top_k
%[\widetilde{\pmb\omega}]_k
%\mathbf{M}_{ck}
%\mathbf{J}_k
%\right)
%\dot{\mathbf{q}},
%\end{equation}
%where $[\widetilde{\pmb\omega}]_k$ is an extended skew-symmetric matrix of $\widetilde{\pmb\omega}$:
%\begin{equation}
%[\widetilde{\pmb\omega}]_k=
%\left[
%\begin{array}{cc}
%\mathbf{0} & \mathbf{0}\\
%\mathbf{0} & \widetilde{\pmb{\omega}}
%\end{array}
%\right]
%=
%\left[
%\begin{array}{cc}
%\mathbf{0} & \mathbf{0}\\
%\mathbf{0} & \widetilde{\mathbf{J}_{\omega k}\dot{\mathbf{q}}}
%\end{array}
%\right].
%\end{equation}
The right-hand side of Eq.~\eqref{eq:eom_rigid} is the generalized external force applied to the skeleton, which includes the gravity force $\mathbf{g}$ and torques $\pmb{\tau}$ from the actuating motors:
\begin{equation}\label{eq:rigid_force}
\mathbf{g}_r=\sum_{k=1}^K\mathbf{J}^\top_k
\left[
\begin{array}{c}
\mathbf{g}_k\\
\mathbf{\pmb\tau}_k
\end{array}
\right]
=
\sum_{k=1}^K\left[\mathbf{J}_{vk}^\top, \mathbf{J}_{\omega k}^\top\right]
\left[
\begin{array}{c}
\mathbf{g}_k\\
\mathbf{\pmb\tau}_k
\end{array}
\right].
\end{equation}

\noindent\textbf{FEM elastic simulation}:
The dynamics of the soft skin can also be formulated using Lagrangian mechanics, and we can obtain the equation of motion in a similar form:
\begin{equation}\label{eq:eom_skin}
\mathbf{M}_d\ddot{\mathbf{u}}+\mathbf{D}_d\dot{\mathbf{u}}+\mathbf{f}_d(\mathbf{u})=\mathbf{g}_d.
\end{equation}
The subscript $[.]_d$ denotes variables for the deformable skin, and $\mathbf{g}_d$ is the external force applied to the soft skin. We discretize the volume of the skin by a tetrahedral mesh. The \emph{deformable mass matrix} $\mathbf{M}_d$ is constant and can be assembled using the standard FEM~\cite{bathe2008finite}. $\mathbf{D}_d$ is the damping matrix, and $\mathbf{f}_d$ denotes the internal elastic force, and it is the negative gradient of the strain energy $\Psi$ such that $\mathbf{f}_d=-\nabla\Psi$. The specific formulation of $\Psi$ depends on the material model chosen. Typically, it is computed based on three \emph{isotropic invariants} of the deformation gradient tensor $\mathbf{F}$:
\begin{equation}\label{eq:invariant}
I_1(\mathbf{F})=\mathtt{tr}(\mathbf{F}^\top\mathbf{F}),\;\; I_2(\mathbf{F})=\mathtt{tr}\left((\mathbf{F}^\top\mathbf{F})^2\right),\;\; I_3(\mathbf{F})=\mathtt{det}(\mathbf{F}^\top\mathbf{F}).
\end{equation}
For the robot with soft skin, its articulated skeleton motion will induce large local compression, especially near the joint. Material models like the StVK and co-rotational models that have been widely used in previous research~\cite{liu2013simulation,capell2002interactive} become unstable under extreme compression. We therefore opted to use the Neo-Hookean material, whose strain energy density is defined as
\begin{equation}\label{eq:neohookean_energy}
\Psi\triangleq\frac{\mu}{2}(I_1-\mathtt{log}(I_3)-3)+\frac{\lambda}{8}\mathtt{log}^2(I_3),
\end{equation}
where $\mu$ and $\lambda$ are Lam\'e constants. The first Piola-Kirchhoff stress tensor $\mathbf{P}\in\mathbb{R}^{3\times3}$ can be computed based on Eq.~\eqref{eq:neohookean_energy} using the chain rule:
\begin{equation}\label{eq:pk1}
\mathbf{P}=\frac{\partial\Psi}{\partial\mathbf{F}}=\frac{\partial\Psi}{\partial I_1}\cdot\frac{\partial I_1}{\partial\mathbf{F}}+\frac{\partial\Psi}{\partial I_3}\cdot\frac{\partial I_3}{\partial\mathbf{F}}=\mu\mathbf{F}-\mu\mathbf{F}^{-\top}+\frac{\lambda\mathtt{log}(I_3)}{2}\mathbf{F}^{-\top}.
\end{equation}
The final formulation of $\mathbf{f}_d$ is
\begin{equation}\label{eq:internal_force}
\mathbf{f}_d=-\int\frac{\partial\Psi}{\partial\mathbf{u}}\mathrm{d}V=-\int\frac{\partial\Psi}{\partial\mathbf{F}}:\frac{\partial\mathbf{F}}{\partial\mathbf{u}}\mathrm{d}V=-\int\left(\mathbf{P}:\frac{\partial \mathbf{F}}{\partial \mathbf{u}}\right)^\top\mathrm{d}V.
\end{equation}
Here, $\partial\mathbf{F}/\partial\mathbf{u}\in\mathbb{R}^{3\times3\times3}$ is a third-order tensor. For a tetrahedral element with linear shape functions,
% $\mathbf{F}$ is invariant within the element and it is linear to the deformed element position. As a result,
$\partial\mathbf{F}/\partial\mathbf{u}$ is constant and can be precomputed and stored at each element.

%\left(\mathbf{M}_r(\mathbf{q}^i)+\Delta t\mathbf{D}_r+ \Delta t^2\mathbf{C}(\mathbf{q}^i, \dot{\mathbf{q}}^i)\right)\Delta\mathbf{q}

\noindent\textbf{Implicit backward euler time integration}:
We discretize the equation of motions for both the rigid skeleton and deformable body using the implicit backward Euler method to improve the stability of the simulation. Let $\Delta \mathbf{q} = \mathbf{q}^{i+1} - \mathbf{q}^{i}$ and  $\Delta \dot{\mathbf{q}} = \dot{\mathbf{q}}^{i+1} - \dot{\mathbf{q}}^{i}$, where the superscript $[\cdot]^{i}$ is the frame index. Given the time interval $\Delta t$ between frame $i$ and $i+1$, the velocity and acceleration of $\mathbf{q}$ at frame $i+1$ can be discretized as: $\dot{\mathbf{q}}^{i+1} = \Delta \mathbf{q}/\Delta t$ and $\ddot{\mathbf{q}}^{i+1} = \Delta \dot{\mathbf{q}}/\Delta t=\Delta \mathbf{q}/\Delta t^2-\dot{\mathbf{q}}^i/\Delta t$. The velocity and acceleration of $\mathbf{u}$ can be derived similarly. Subsequently, Eqs.~\eqref{eq:eom_rigid} and \eqref{eq:eom_skin} can be linearized as
\begin{multline}\label{eq:eom_rigid_discretization}
\left(\mathbf{M}_r(\mathbf{q}^i)+\Delta t\mathbf{D}_r+ \mathbf{C}(\mathbf{q}^i, \dot{\mathbf{q}}^i)\right)\Delta\mathbf{q} = \Delta t^2\big(\mathbf{g}_r - \mathbf{D}_r\dot{\mathbf{q}}^i\big),
\end{multline}
and
\begin{equation} \label{eq:eom_soft_discretization}
\left(\mathbf{M}_d + {\Delta t}\mathbf{D}_d + {\Delta t^2}\frac{\partial \mathbf{f}_d}{\partial \mathbf{u}^i} \right)\Delta \mathbf{u}
 = {\Delta t}^2\big(\mathbf{g}_d - \mathbf{f}_d(\mathbf{u}^i) - \mathbf{D}_d\dot{\mathbf{u}}^i\big).
\end{equation}
In Eq.~\eqref{eq:eom_rigid_discretization}, we compute $\mathbf{M}_r(\mathbf{q}^i)$ and $\mathbf{C}(\mathbf{q}^i, \dot{\mathbf{q}}^i)$ using state variables at frame $i$. Therefore, Eq.~\eqref{eq:eom_rigid} is only semi-implicitly discretized~\cite{Baraff:1998,Peng2004}.

\noindent\textbf{Linear systems}: The two-way coupled multibody-elastic system in Eq.[2] in Sec. 4.1 of our paper is as follows:
\begin{equation}
\left[
\begin{array}{ccc}
\mathbf{A}_r  & \mathbf{0} & \nabla_q\mathcal{C}^\top\\
\mathbf{0}    & \mathbf{A}_d & \nabla_u\mathcal{C}^\top\\
\nabla_q\mathcal{C} & \nabla_u\mathcal{C} & \mathbf{0}
\end{array}
\right]
\left[
\begin{array}{c}
\Delta\mathbf{q}\\
\Delta\mathbf{u}\\
\pmb\lambda
\end{array}
\right]
=
\left[
\begin{array}{c}
\mathbf{b}_r\\
\mathbf{b}_d\\
\mathbf{0}
\end{array}
\right].
\label{eq:system_equation}
\end{equation}

The matrices $\mathbf{A}_r, \mathbf{A}_d$ in this equation are just the abbreviation of system matrices in Eqs.~\eqref{eq:eom_rigid_discretization} and \eqref{eq:eom_soft_discretization}, where $\mathbf{A}_r = \mathbf{M}_r(\mathbf{q}^i)+\Delta t\mathbf{D}_r+ \Delta t^2\mathbf{C}(\mathbf{q}^i, \dot{\mathbf{q}}^i)$ and $\mathbf{A}_d = \mathbf{M}_d + {\Delta t}\mathbf{D}_d + {\Delta t^2} {\partial \mathbf{f}_d}/{\partial \mathbf{u}^i}$. Analogically, we have $\mathbf{b}_r = \Delta t^2\big(\mathbf{g}_r - \mathbf{D}_r\dot{\mathbf{q}}^i -  \mathbf{C}(\mathbf{q}^i, \dot{\mathbf{q}}^i)\dot{\mathbf{q}}^i\big)$, and $\mathbf{b}_d = \Delta t^2\big(\mathbf{g}_d - \mathbf{f}_d(\mathbf{u}^i) - \mathbf{D}_d\dot{\mathbf{u}}^i\big)$.

\section{System Matrix Condensation}
The system matrix condensation in Sec.5.2 starts with eliminating the Lagrange multipliers $\mathbf{\lambda}$ in Eq.~\eqref{eq:system_equation} (the same equation with Eq.[2] in Sec. 4.1). We first expand the second line of this equation, which yields:
\begin{equation}\label{eq:condensation_u}
\mathbf{A}_d\Delta\mathbf{u}+\nabla_u\mathcal{C}^\top\pmb{\lambda}=\mathbf{b}_d,\quad\text{or}\quad\Delta\mathbf{u}=\mathbf{A}_d^{-1}\big(\mathbf{b}_d-\nabla_u\mathcal{C}^\top\pmb{\lambda}\big).
\end{equation}
Similarly, expanding the first line in Eq.~\eqref{eq:system_equation}, we can produce an equation similar to Eq.~\eqref{eq:condensation_u} but for skeleton DOFs:
\begin{equation}\label{eq:condensation_q}
\Delta\mathbf{q}=\mathbf{A}_r^{-1}\left(\mathbf{b}_r-\nabla_q\mathcal{C}^\top\pmb{\lambda}\right).
\end{equation}
Expanding the third line of Eq.~\eqref{eq:system_equation} gives the linearized position constraint:
\begin{equation}\label{eq:condensation_lambda}
\nabla_q\mathcal{C}\Delta\mathbf{q}+\nabla_u\mathcal{C}\Delta\mathbf{u}=\mathbf{0}.
\end{equation}
Substituting both Eqs.~\eqref{eq:condensation_u} and \eqref{eq:condensation_q} into Eq.~\eqref{eq:condensation_lambda} yields:
%\begin{equation}\label{eq:condensation_lambda3}
%\nabla_q\mathcal{C}\mathbf{A}_r^{-1}\left(\mathbf{b}_r-\nabla_q\mathcal{C}^\top\pmb{\lambda}\right)+\nabla_u\mathcal{C}\mathbf{A}_d^{-1}\left(\mathbf{b}_d-\nabla_u\mathcal{C}^\top\pmb{\lambda}\right)=\mathbf{0}.
%\end{equation}
%It leads to:
\begin{equation}\label{eq:condensation_lambda4}
\pmb{\lambda}=\mathbf{A}_C^{-1}\left(\nabla_q\mathcal{C}\mathbf{A}_r^{-1}\mathbf{b}_r+\nabla_u\mathcal{C}\mathbf{A}_d^{-1}\mathbf{b}_d\right),
\end{equation}
where $\mathbf{A}_C=\nabla_q\mathcal{C}\mathbf{A}_r^{-1}\nabla_q\mathcal{C}^\top+\nabla_u\mathcal{C}\mathbf{A}_d^{-1}\nabla_u\mathcal{C}^\top$.
By substituting Eq.~\eqref{eq:condensation_lambda4} back into Eqs.~\eqref{eq:condensation_u} and \eqref{eq:condensation_q}, we obtain the condensed formulas in Eq.[10] in our paper.
\bibliographystyle{IEEEtran}
\bibliography{softrobot}
%\begin{thebibliography}{1}
%H.~Kopka and P.~W. Daly, \emph{A Guide to {\LaTeX}}, 3rd~ed.\hskip 1em plus
%\end{thebibliography}

\vspace{-60pt}
\begin{IEEEbiography}[{\includegraphics[width=1in,height=1.25in,clip,keepaspectratio]{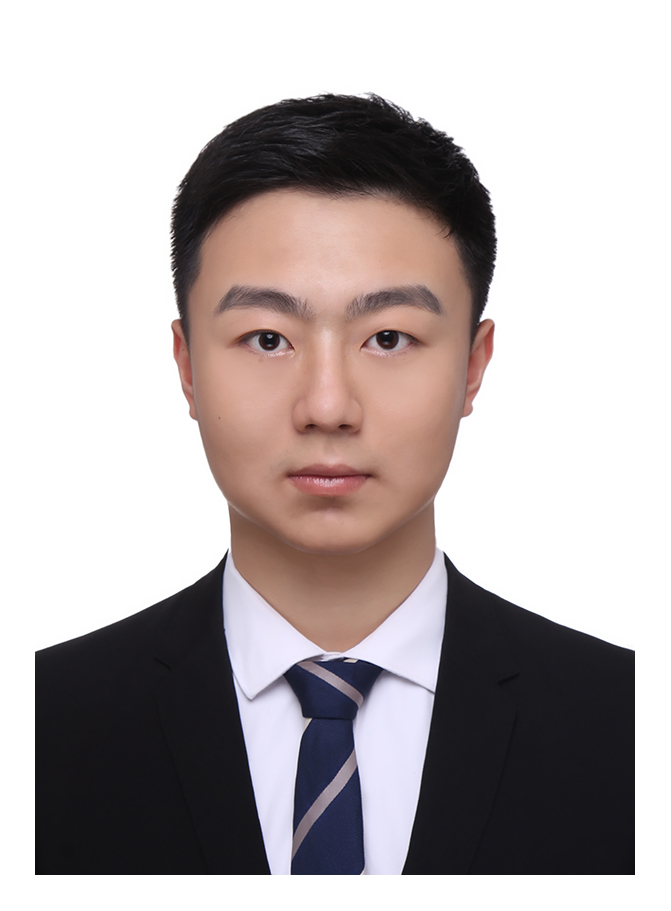}}]{Xudong Feng} Xudong Feng is a Ph.D. candidate in the State Key Lab of CAD \& CG, Colledge of Computer Science, Zhejiang University. He received his bachelor degree in Tianjin University. His research interests include physical based simulation and animation, digital fabrication and reinforcement learning.
\end{IEEEbiography}

\vspace{-10pt}
\begin{IEEEbiography}[{\includegraphics[width=1in,height=1.25in,clip,keepaspectratio]{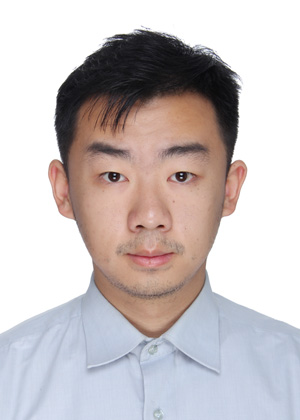}}]{Jiafeng Liu} is currently working toward the Master degree in computer engineering with the Zhejiang University. He received the bachelor's degree from the Hefei University of Technology in 2018. He  His research interests include character animation, robotics locomotion, machine learning and related topics.
\end{IEEEbiography}

\vspace{-60pt}
\begin{IEEEbiography}[{\includegraphics[width=1in,height=1.25in,clip,keepaspectratio]{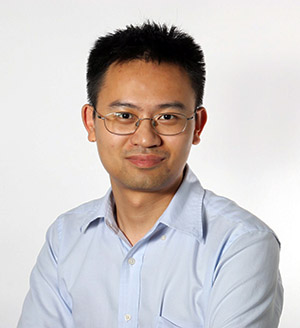}}]{Huamin Wang} received the BEng degree from
Zhejiang University, the MS degree from Stanford
University, and the PhD degree in computer science
from the Gerogia Institute of Technology, in
2002, 2004 and 2009. Hw is an associate professor
in the Department of Computer Science and
Engineering, the Ohio State University. Before joining
Ohio State University, he was a postdoctoral
researcher in the Department of Electrical Engineering
and Computer Sciences, the University of
California, Berkeley. He is a member of the IEEE.
\end{IEEEbiography}

\vspace{-60pt}
\begin{IEEEbiography}[{\includegraphics[width=1in,height=1.25in,clip,keepaspectratio]{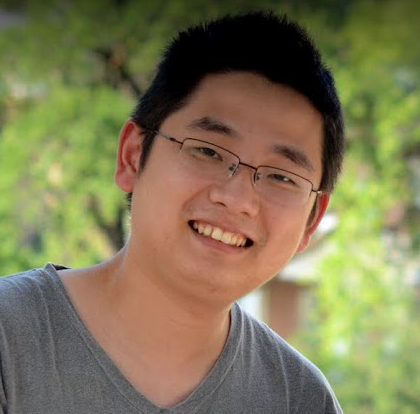}}]{Yin Yang} received the PhD degree in computer
science from the University of Texas at Dallas, in
2013. He is an assistant professor in the Department
of Electrical Computer Engineering, the
University of New Mexico, Albuquerque. His
research interests include physics-based animation/
simulation and related applications, scientific
visualization, and medical imaging analysis. He is
a member of the IEEE.
\end{IEEEbiography}

%\newpage

\vspace{-60pt}
\begin{IEEEbiography}[{\includegraphics[width=1in,height=1.25in,clip,keepaspectratio]{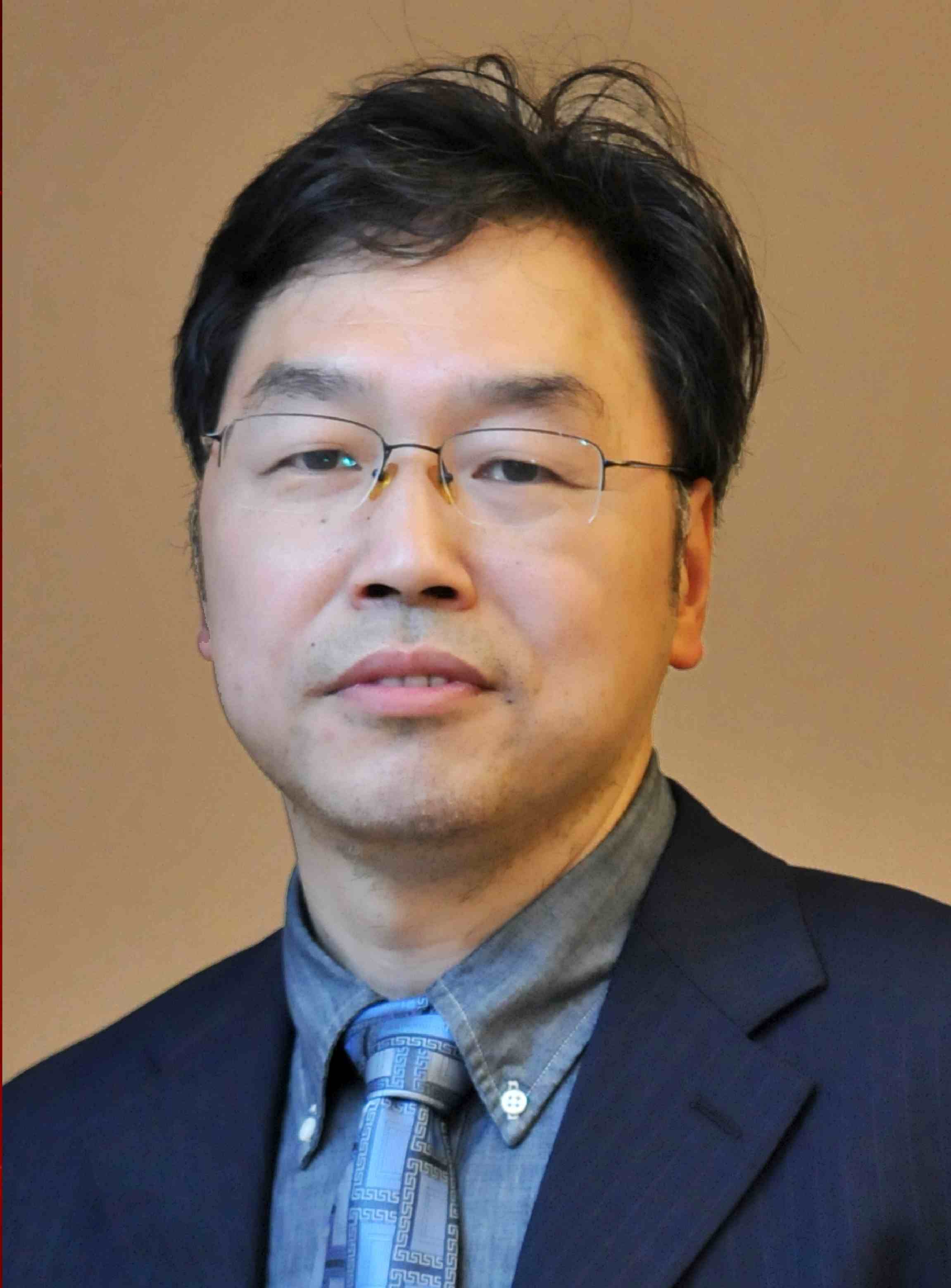}}] {Hujun Bao} is a Chenkong professor in State Key Lab of
CAD\&CG, College of Computer Science at Zhejiang University. He received PhD degree in applied mathematics from Zhejiang university in 1993. His
main research interests are computer graphics and computer vision, including
real-time rendering technique, geometry computing, virtual reality,
and 3D reconstruction, and has published more than 100 papers on prestigious academic journals and international conferences.
\end{IEEEbiography}

\vspace{-60pt}
\begin{IEEEbiography}[{\includegraphics[width=1in,height=1.25in,clip,keepaspectratio]{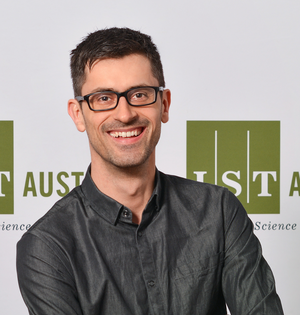}}]{Bernd Bickel}
 is an Assistant Professor, heading the Computer Graphics and Digital Fabrication group at IST Austria. He received Master and Ph.D degree in Computer Science from ETH Zurich. His research interests includeds computer graphics and its overlap into robotics, computer vision, biomechanics, material science, and digital fabrication. He receives SIGGRPAH significant researcher award in 2017.
\end{IEEEbiography}

\vspace{-60pt}
\begin{IEEEbiography}[{\includegraphics[width=1in,height=1.25in,clip,keepaspectratio]{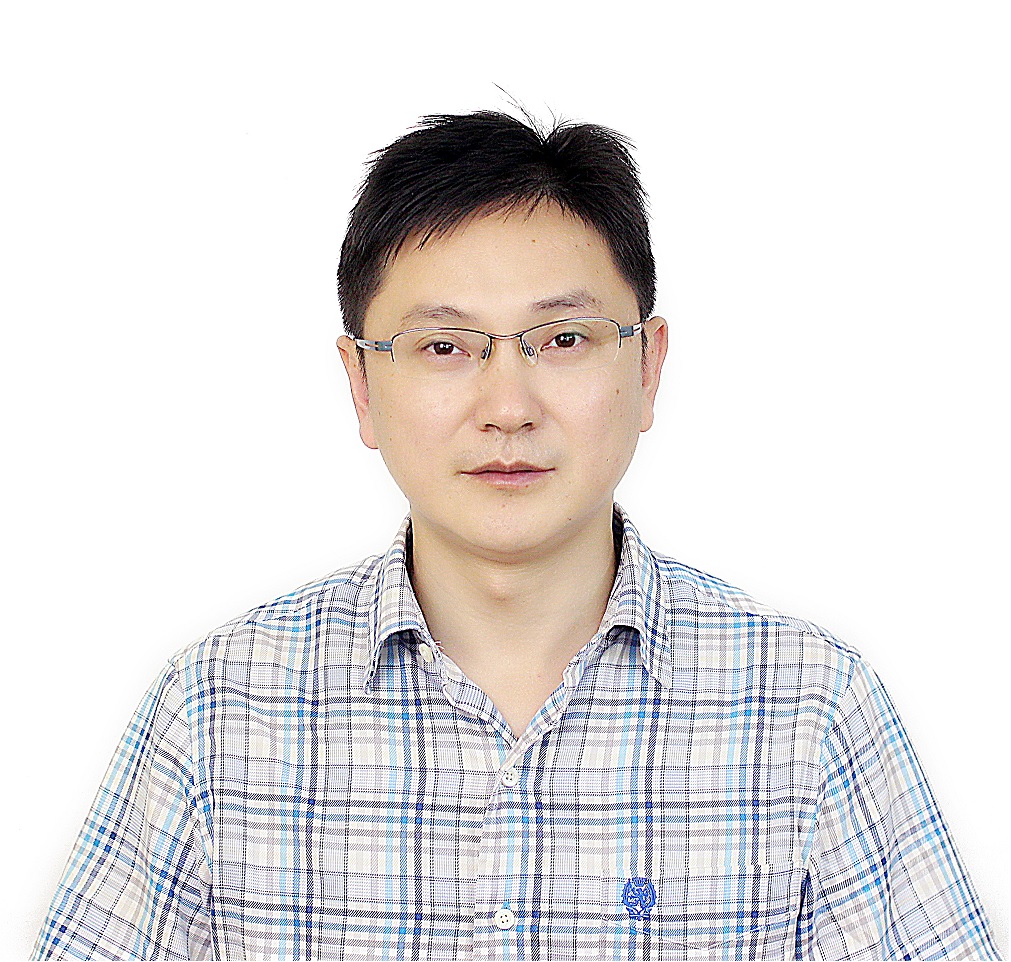}}]{Weiwei Xu} is a researcher with the State Key Lab
of CAD \& CG, College of Computer Science,
Zhejiang University, awardee of NSFC Excellent
Young Scholars Program in 2013. His main
research interests include the digital geometry
processing, physical simulation, computer vision and virtual reality.
He has published around 70 papers on international
graphics journals and conferences, including
16 papers on ACM TOG. He is a member of
the IEEE.
\end{IEEEbiography}


%\vfill


\end{document}